\documentclass[a4paper,fleqn,usenatbib]{mnras}

\usepackage[T1]{fontenc}
\usepackage{ae,aecompl}

\usepackage{graphicx}	
\usepackage{amsmath}	
\usepackage{amssymb}	
\usepackage{multicol}        
\usepackage{bm}
\usepackage{xcolor}
\usepackage{soul}		

\title[Rotational modulation in {\it TESS} B stars]
{Rotational modulation in {\it TESS} B stars}
\author[L.A. Balona et al.]
{L. A. Balona$^1$, G. Handler$^2$, S. Chowdhury$^2$, D. Ozuyar$^3$,
C. A. Engelbrecht$^4$, \newauthor
G. M. Mirouh$^5$, G. A. Wade$^6$, A. David-Uraz$^7$, M. Cantiello$^{8,9}$\\
$^1$South African Astronomical Observatory, P.O. Box 9, Observatory, Cape
Town 4735, South Africa\\
$^2$Nicolaus Copernicus Astronomical Center, Bartycka 18, PL-00-716 Warsaw,
Poland\\
$^3$Department of Astronomy and Space Science, Ankara University, 06100
Tandogan-Ankara, Turkey\\
$^4$Department of Physics, University of Johannesburg, PO Box 524, Auckland
Park, Johannesburg, South Africa\\
$^5$Astrophysics Research Group, Faculty of Engineering and Physical Sciences, 
University of Surrey, Guildford GU2 7XH, UK\\
$^6$Department of Physics \& Space Science, Royal Military College of Canada,
P.O. Box 17000, Station Forces,\\
Kingston, Ontario, Canada, K7K 7B4\\
$^7$Department of Physics and Astronomy, University of Delaware, Newark,
DE 19716, USA\\
$^8$Center for Computational Astrophysics, Flatiron Institute, 162
5th Avenue, New York, NY 10010, USA\\
$^9$Department of Astrophysical Sciences, Princeton University,
Princeton, NJ 08544, USA\\
}

\begin{document}

\date{Accepted .... Received ...}

\pagerange{\pageref{firstpage}--\pageref{lastpage}} \pubyear{2011}

\maketitle

\label{firstpage}

\begin{abstract}
Light curves and periodograms of 160 B stars observed by the {\it TESS} space 
mission and 29 main-sequence B stars from {\it Kepler} and {\it K2} 
were used to classify the variability type.  There are 114 main-sequence B stars in
the {\it TESS} sample, of which 45 are classified as possible rotational 
variables.  This confirms previous findings that a large fraction (about 
40\,percent) of A and B stars may exhibit rotational modulation.  {\it Gaia} 
DR2 parallaxes were used to estimate luminosities, from which the radii and 
equatorial rotational velocities can be deduced.  It is shown that observed 
values of the projected rotational velocities are lower than the estimated 
equatorial velocities for nearly all the stars, as they should be if rotation 
is the cause of the light variation.  We conclude that a large fraction of 
main-sequence B stars appear to contain surface features which cannot likely 
be attributed to abundance patches.
\end{abstract}

\begin{keywords}
stars: early-type - stars:rotation - stars: starspots - stars:oscillations 
\end{keywords}

\section{Introduction}

The existence of large spots or spot groups on the surfaces of cool stars
other than the Sun is well established.  This discovery can be traced 
back to \citet{Kron1947} who observed four eclipsing binaries and detected 
significant light variability outside eclipse that could not be explained 
other than by the presence of spots similar to those on the Sun. These stars 
were later called RS~CVn binaries. \citet{Hall1972} was the first to 
explicitly postulate the starspot model in these stars.  Rotational modulation
due to starspots is also detected in the BY~Dra variables, which are 
emission-line K and M dwarfs, and in the FK~Com stars, which are rapidly 
rotating G-K giants with emission in the CaII lines.  Over 500 field stars 
showing evidence of starspots are known (see \citealt{Strassmeier2009} for a 
review), but many thousands have been detected from the {\it Kepler} space 
mission \citep{McQuillan2013,McQuillan2014,Reinhold2013,Nielsen2013,
Chowdhury2018}.  

Sunspots appear cooler than the surrounding photosphere because they
correspond to regions of lower convective energy transport.  The decrease in
energy transport is due to strong localized magnetic fields which affect the
convective motions close to the stellar surface.  The magnetic fields are
thought to be a result of dynamo action in the convective outer envelope of 
the Sun and other cool stars \citep{Charbonneau2014}.  From this perspective, 
only stars with outer convective envelopes can support such a magnetic field.  
Hence dark starspots are not expected in A and B stars which have radiative 
envelopes.

The chemically peculiar Ap and Bp stars do, however, show rotational light 
modulation due to patches of differing chemical abundances on the stellar
surface.  The chemical peculiarities are believed to be confined to the outer 
layers of the star and are generally thought to be a result of gravitational 
settling and diffusion of elements in the presence of a strong global 
magnetic field \citep{Michaud1970}. For these stars, which have radiative 
atmospheres, the magnetic field is thought to be of fossil origin.

The discovery that a large fraction of A and B stars observed by {\it Kepler} 
seem to show rotational modulation \citep{Balona2013c,Balona2016a,Balona2017a} 
and some of them possibly even flares \citep{Balona2012c} was therefore
unexpected. The possible existence of rotational modulation among the B stars 
suggests that our current understanding of the physics of the outer 
layers of hot stars may need to be revised.  The fact that a great majority of
$\delta$~Scuti stars show unexpected low frequencies \citep{Balona2018c},
which cannot be explained by current models, also points to such a revision.

The Transiting Exoplanet Survey Satellite mission ({\it TESS}; \citealt{Ricker2015}) 
is designed to search for exoplanets. A preliminary report on B stars from
the first set of observations (Sectors 1 and 2) covering 55\,d can be found in
\citet{Pedersen2019}.  Our main aim is to classify the {\it TESS} B stars 
observed in Sectors 1 and 2 into various variability types, to estimate the 
approximate fraction of main sequence stars which might be rotational 
variables and to determine if the photometric periods are consistent with the 
presumed rotational periods.

\section{Data and analysis}

{\it TESS} observes the sky in sectors measuring $24^\circ \times 96^\circ$ that 
extend from near the ecliptic equator to beyond the ecliptic pole. 
Each sector is observed for two orbits of the satellite around the Earth, 
or about 27 days.  Sectors begin to overlap towards the ecliptic pole which
means that at mid-latitudes the same star will be observed in more than one
sector.  There is a continuous viewing zone near the ecliptic pole where
the same stars are observed in all sectors.  

The data analyzed in this paper are from the light curves of the first release 
(Sector 1 and Sector 2). Most of the stars discussed here are at mid-latitudes 
and have been observed for a time span of about 55\,d.  A description of how 
these data products were generated is found in \citet{Jenkins2016}.  

Light curves are generated with two-minute cadence using simple aperture 
photometry (SAP) and pre-search data conditioning (PDC).  The PDC pipeline 
module uses singular value decomposition to identify and correct for 
time-correlated instrumental signatures in the light curves.  In addition,
PDC corrects the flux for each target to account for crowding from other 
stars and their effects.  Only PDC light curves are used in this paper. 

The noise level in the amplitude periodogram is around 10\,ppm for the 
brightest stars, about 30\,ppm at about $V = 8$ mag, 100\,ppm at $V = 10$\,mag 
and 200\,ppm at $V = 12$\,mag.  A frequency peak with a false alarm probability
of $10^{-3}$ or less is taken as being significant.

The {\it TESS} input catalogue \citep{Stassun2018} lists stellar parameters for
stars observed by {\it TESS}.  The effective temperatures, $T_{\rm eff}$,
for stars without spectroscopic determinations were obtained from
near-infrared photometry, which is not reliable for B stars, particularly
since reddening is important in may cases.  The stars observed
by {\it TESS} were matched with the SIMBAD astronomical database
\citep{Wenger2000} and only those known to be B stars were selected,
giving a total of 160 stars with spectral types earlier than A0.

Periodograms and light curves were visually inspected and each star assigned 
a variability class where appropriate.  A necessary  signature of rotational 
modulation is taken to be a significant, isolated low-amplitude peak with 
frequency less than 4\,d$^{-1}$ or a low-amplitude peak with one or more 
harmonics.  The classification was made independently by LAB and GH with 
agreement among the stars deemed to be rotational variables.  

\section{Rotational variability}

The equatorial rotational velocity, $v_e$ (km\,s$^{-1}$), is given by 
$v_e = 50.74\nu_{\rm rot}(R/R_\odot)$, where the rotational frequency, 
$\nu_{\rm rot}$ is in cycles d$^{-1}$ and $R/R_\odot$ is the stellar radius 
in solar units.  For main-sequence B stars the radii are typically 
2--10 $R/R_\odot$ and $v_e$ in the range 0--400\,km\,s$^{-1}$.  Thus one 
might expect $0 < \nu_{\rm rot} < 4$\,d$^{-1}$.  A periodogram peak in this
frequency range could be a result of rotational modulation.  

There are two ways to show that the variability of a group of stars might be 
due to rotational modulation.  One way is to demonstrate that there is a
relationship between the projected rotational velocities, $v \sin i$, 
and the equatorial rotational velocities, $v_e$. Since $\sin i \leq 1$, the 
expectation is that in a plot of $v\sin i$ as a function of $v_e$, the points 
will all lie on or below the line $v\sin i = v_e$, subject to measurement 
errors (see Fig.\,2 in \citealt{Balona2017a}).  

Another method is to show that, for stars in the main sequence band, the
distribution of $v_e$, derived from $\nu_{\rm rot}$ and an estimate of the
stellar radii, matches the distribution of $v_e$ derived from spectroscopic
measurements of $v\sin i$ for stars in the general field within the same 
temperature range (see Fig.\,8 in \citealt{Balona2013c}).  This has the 
advantage that values of $v\sin i$ of the stars to be tested are not required.
A sufficient number of stars in each $v_e$ bin is necessary and the number of 
bins needs to be sufficiently large to adequately resolve the distribution. 
The method therefore requires a rather large number of stars.  The number of 
B stars with photometrically derived rotation periods is, at present, too few 
for this method to be applied. 

Classification of stars according to variability type is an essential first
step in any analysis.  The information at hand is very limited: the light 
curve, the periodogram and the approximate location of the star in the 
H-R diagram as judged by the spectral type and the {\it Gaia} DR2 parallax.  

It is reasonable to adopt the variability type definitions in the 
{\it General Catalogue of  Variable Stars} (GCVS, \citealt{Samus2009}).  
There may be variability which does not seem to fit in any of the GCVS  
classes.  Unless there is additional supporting evidence, the temptation must 
be resisted to assign a new class of variable.  Since the GCVS does not 
include a type for rotational variables among the A and B stars, we have 
chosen ROT as a suitable designation for any star exhibiting rotational 
modulation, but not known to be chemically peculiar.

The SXARI variables are a specific set of B0p-B9p rotational variables with 
variable-intensity HeI and SiIII lines and magnetic fields.  The periods of 
their light and magnetic field variations are consistent with rotation.  They 
are high-temperature analogs of the $\alpha^2$~CVn (ACV) variables, which are Ap 
stars with a tilted global magnetic field and abundance patches, giving 
rise to rotational modulation of the light curve.  If a star shows signs of 
abnormalities, such as enhanced Si, Mg or Hg or is He weak or He strong, then 
the variability is likely a result of  a patch or patches of enhanced chemical
abundance.  For the purpose of this study, we classify these stars as SXARI 
rather than ACV or ROT so as to distinguish between two different causes of
rotational modulation: localized abundance anomalies or localized
temperature differences.  The photometric frequency in both cases is the
rotational frequency.
  
Binaries in a circular orbit may give rise to a low frequency peak either
through tidal effects (ELL type of variable) or grazing eclipses. Eclipses 
are easy to spot as they give rise to a large number of harmonics in the 
periodogram.  Tidal effects cannot easily be distinguished from rotational 
modulation. The ambiguity between the ELL and ROT classifications can be 
broken if there is significant amplitude or frequency modulation.

\begin{table*}
\begin{center}
\caption{List of B stars observed by {\it TESS}.  The TIC number and the star 
name is given in the first two columns.  This is followed by the  assigned 
variability type (GCVS type in brackets) and the apparent $V$ magnitude. 
The presumed rotational frequency, $\nu_{\rm rot}$, the amplitude,
$A_{\rm rot}$, the signal-to-noise ratio, S/N, of the periodogram peak and
the number of visible harmonics, $N_H$, is shown. The derived equatorial 
rotational velocity is $v_e$ and $v\sin i$ is the projected rotational velocity 
from the literature.  The adopted effective temperature is $T_{\rm eff}$ as
derived from the reference (see Table\,{\ref{refs}}).  Finally, the stellar 
luminosity obtained from the {\it Gaia} parallax and the spectral type 
are shown (``(Be)'' indicates a classical Be star).}
\label{bstars}
\resizebox{18cm}{!}{
\begin{tabular}{rllrrrrrrrrrrl}
\hline
\multicolumn{1}{c}{TIC}               & 
\multicolumn{1}{l}{Name}              & 
\multicolumn{1}{l}{Var. Type}         & 
\multicolumn{1}{c}{$V$}               & 
\multicolumn{1}{c}{$\nu_{\rm rot}$}   & 
\multicolumn{1}{c}{$A_{\rm rot}$}     & 
\multicolumn{1}{c}{S/N}     & 
\multicolumn{1}{c}{$N_H$}             & 
\multicolumn{1}{c}{$v_e$}             & 
\multicolumn{1}{c}{$v\sin i$}         & 
\multicolumn{1}{c}{$T_{\rm eff}$}     & 
\multicolumn{1}{c}{Ref}               & 
\multicolumn{1}{c}{$\log\tfrac{L}{L_\odot}$} & 
\multicolumn{1}{l}{Sp. Type}          \\
\multicolumn{1}{c}{}                  & 
\multicolumn{1}{l}{}                  & 
\multicolumn{1}{l}{}                  & 
\multicolumn{1}{c}{mag}               &
\multicolumn{1}{c}{(d$^{-1}$)}        & 
\multicolumn{1}{c}{(ppm)}             & 
\multicolumn{1}{c}{}                  & 
\multicolumn{1}{c}{}                  & 
\multicolumn{1}{c}{(km/s)}            & 
\multicolumn{1}{c}{(km/s)}            & 
\multicolumn{1}{c}{(K)}               &
\multicolumn{1}{c}{}                  &
\multicolumn{1}{c}{}                  &
\multicolumn{1}{l}{}                 \\
\hline
  12359289 & HD 225119         & SXARI           &  8.180  & 0.325 &  4157 &  95 & 1 &   65 &      &  15330 &  1 & 2.90 & kB8hB7HeB9.5IIISi \\
  29990592 & HD 268623         & ACYG            & 11.635  &       &       &     &   &      &      &  20665 &  2 &      & B1.5Ia (LMC) \\
  30110048 & HD 268653         & ACYG            & 10.760  &       &       &     &   &      &      &  17185 &  2 &      & B2.5Ia (LMC) \\
  30268695 & HD 268809         & ACYG            & 11.964  &       &       &     &   &      &      &  22845 &  2 &      & B0.5Ia (LMC) \\
  30275662 & Sk-66 27          & ACYG            & 11.779  &       &       &     &   &      &      &  17765 &  2 &      & B2.5/3Ia (LMC) \\
  30312676 & HD 268726         & ACYG            & 11.265  &       &       &     &   &      &      &  19075 &  2 &      & B2Iaq (LMC) \\
  30317301 & HD 268798         & EB              & 11.490  &       &       &     &   &      &      &  28000 &  1 &      & B0.5, B2Ia (LMC) \\
  30933383 & Sk-68 39          & ACYG            & 12.039  &       &       &     &   &      &      &  22580 &  1 &      & B2.5Ia (LMC) \\
  31105740 & TYC 9161-925-1    & ACYG            & 12.010  &       &       &     &   &      &      &  24970 &  2 &      & B0.5Iae (LMC) \\
  31181554 & HD 269050         & ACYG            & 11.540  &       &       &     &   &      &      &  25000 &  1 &      & B0Ia(e?) (LMC) \\
  31674330 & GJ 127.1          & -               & 11.394  &       &       &     &   &      &    0 &  16860 &  3 &-2.75 & DA3.0 \\
  31867144 & HD 22252          & SPB             &  5.806  &       &       &     &   &      &  223 &  12157 &  4 & 2.72 & B8IV \\
  33945685 & HD 223118         & ROT             &  8.250  & 2.835 &    72 &  25 & 0 &  273 &      &  10500 &  5 & 1.60 & B9V  \\
  38602305 & HD 27657          & ROT             &  5.870  & 0.336 &   920 & 143 & 3 &   42 &      &  12448 &  7 & 2.13 & B9III/IV; A2-A7m \\
  40343782 & HD 269101         & ACYG?           & 12.030  &       &       &     &   &      &      &  21370 &  1 &      & B3Iab; (LMC) \\
  41331819 & HD 43107          & ROT             &  5.044  & 0.714 &   324 &  91 & 0 &  106 &   98 &  10886 &  4 & 2.04 & B9.5III, B8V \\
  47296054 & HD 214748         & SPB+ROT         &  4.180  & 0.836 &    63 &  51 & 3 &  202 &  180 &  13520 &  1 & 2.84 & B8IVe (Be) \\
  49687057 & HD 220787         & -               &  8.290  &       &       &     &   &      &   26 &  17379 &  6 & 3.54 & B3III \\
  53992511 & HD 209522         & BE (BE)         &  5.952  &       &       &     &   &      &  280 &  22570 &  1 & 3.51 & B4IVe (Be) \\
  55295028 & HD 33599          & BE              &  8.970  &       &       &     &   &      &  200 &  22570 &  1 & 4.44 & B3p shell: (Be) \\
  66497441 & HD 222847         & ELL             &  5.235  &       &       &     &   &      &  307 &  12000 &  8 & 2.24 & B8.5Vnn \\
  69925250 & V* HN Aqr         & BCEP+SPB (BCEP) & 11.470  &       &       &     &   &      &   45 &  22909 &  8 & 4.31 & B0/1 \\
  89545031 & HD 223640         & SXARI (ACV)     &  5.180  & 0.266 &  9000 &  95 & 1 &   36 &   27 &  12462 &  9 & 2.20 & B9SiSrCr* \\
  92136299 & HD 222661         & ROT+FLARE?      &  4.483  & 2.251 &   217 &  82 & 1 &  225 &  120 &  11108 &  8 & 1.73 & B9.5IV \\
 115177591 & HD 201108         & ROT             &  6.900  & 1.736 &  2442 &  89 & 0 &  345 &      &  10158 & 10 & 2.17 & B8IV/V \\
 118327563 & CD-38 222         & ROT+FLARE?      & 10.260  & 4.372 &   733 &  60 & 1 &   54 &   48 &  26300 & 11 & 1.42 & sdB \\  
 139468902 & HD 213155         & ROT             &  6.924  & 2.199 &   288 &  87 & 1 &  264 &      &   9628 &  6 & 1.64 & B9.5V  \\
 141281495 & HD 37854          & ROT             &  8.100  & 0.334 &   205 &  89 & 0 &   48 &      &  10500 &  5 & 1.96 & B9/9.5V \\
 147283842 & HD 205805         & -               & 10.180  &       &       &     &   &      &      &  25000 & 11 & 1.67 & sdB4 \\
 149039372 & HD 34543          & SPB?            &  8.370  &       &       &     &   &      &      &  11583 &  6 & 2.12 & B8V \\
 149971754 & HD 41297          & SPB             & 10.000  &       &       &     &   &      &      &  13520 &  7 & 2.57 & B8IV \\
 150357404 & HD 45796          & SPB+ROT?        &  6.248  & 0.640 & 10333 &  77 & 0 &  114 &   17 &  13775 &  6 & 2.61 & B6V \\
 150442264 & HD 46792          & EB (EB:)        &  6.140  &       &       &     &   &      &      &  16605 &  6 & 3.42 & B3(V)k \\
 152283270 & HD 208433         & EA+ROT?         &  7.440  & 0.434 &    80 &  26 & 0 &   69 &      &  10500 &  5 & 2.04 & B9.5V \\
 167045028 & HD 45527          & EB?             &  9.910  & 0.329 &  5811 & 135 & 1 &   62 &      &  11000 &  1 & 2.27 & B9IV \\
 167415960 & HD 48467          & -               &  8.270  &       &       &     &   &      &      &  10613 &  7 & 2.19 & B8/9V \\
 167523976 & HD 49193          & SPB             &  8.940  &       &       &     &   &      &      &  24380 &  1 & 4.06 & B2V \\
 169285097 & CD-35 15910       & sdB Hybrid      & 11.000  &       &       &     &   &      &      &  28390 & 11 & 1.50 & sdB He1 \\
 176935619 & HD 49306          & ROT             &  6.700  & 2.748 &    29 &  26 & 0 &  326 &      &  10082 &  7 & 1.71 & B9.5/A0V \\
 176955379 & HD 49531          & ROT             &  8.910  & 2.994 &  1181 & 109 & 1 &  414 &      &  11900 &  5 & 2.13 & B8/9Vn \\
 177075997 & HD 51557          & ROT?            &  5.393  & 0.352 &    49 &  44 & 0 &   89 &  124 &  12325 &  6 & 2.72 & B7III \\
 179308923 & HD 269382         & ACYG+EB?        & 10.815  &       &       &     &   &      &      &  27600 &  1 &      & O9.5Ib (LMC) \\
 179574710 & HD 271213         & ACYG            & 12.310  &       &       &     &   &      &      &  23790 &  1 &      & B1Iak (LMC)\\
 179637387 & [OM95]LH47-373A   & ACYG?           & 11.970  &       &       &     &   &      &      &  20300 & 12 &      & B1Ib (LMC)\\
 179639066 & HD 269440         & ACYG            & 11.378  &       &       &     &   &      &      &  22240 &  2 &      & B1.5Ia (LMC) \\
 181043970 & HD 5148           & EA (EA/SD)      & 10.640  &       &       &     &   &      &      &  11000 &  5 & 1.41 & B9/A2IV: \\
 182909257 & HD 6783           & SXARI           &  7.940  & 0.319 &  2837 &  92 & 3 &   36 &   30 &  12941 &  6 & 2.11 & B8Si \\
 197641601 & HD 207971         & ROT?            &  3.010  & 0.201 &   135 &  36 & 0 &   68 &   55 &  11984 &  6 & 2.92 & B8IV-Vs \\
 206362352 & HD 223145         & SPB             &  5.161  &       &       &     &   &      &  240 &  17163 &  4 & 3.03 & B2.5V \\
 206547467 & HD 210780         & ROT             &  8.340  & 0.233 &    75 &  25 & 1 &   16 &      &  12500 &  7 & 1.61 & B9.5/A0V \\
 207176480 & HD 19818          & BE              &  9.060  &       &       &     &   &      &      &  11710 &  5 & 1.59 & B9/A0Vne: (Be) \\
 207235278 & HD 20784          & ELL             &  8.280  & 0.558 &  4530 &  95 & 1 &  187 &      &   9175 &  7 & 2.45 & B9.5V \\
 220430912 & HD 31407          & EA (EA)         &  7.690  &       &       &     &   &      &      &  19648 &  7 & 3.76 & B2/3V \\
 224244458 & HD 221507         & SXARI+FLARE     &  4.370  & 0.522 &   141 &  80 & 2 &   57 &   21 &  12380 &  6 & 2.00 & B9.5IVpHg:Mn:Eu:\\
 229013861 & HD 208674         & ROT?            &  7.920  & 0.423 &    80 &  23 & 0 &   35 &      &  14555 &  7 & 2.04 & B9.5V \\
 230981971 & HD 10144          & BE (BE)         &  0.460  &       &       &     &   &      &  225 &  20760 &  1 &        & B4V(e) (Be) \\
 231122278 & HD 29994          & SPB             &  8.110  &       &       &     &   &      &      &  11900 &  5 & 2.08 & B8/9V \\
 238194921 & HD 24579          & SPB+ROT         &  8.060  & 0.727 &    74 &  37 & 0 &  114 &      &  15330 &  1 & 2.68 & B7III \\
 259862349 & HD 16978          & ROT?            &  4.106  & 0.402 &    51 &  31 & 0 &   53 &   96 &  10003 &  6 & 1.79 & B9V  \\
 260128701 & HD 42918          & SPB             &  7.950  &       &       &     &   &      &      &  16289 &  7 & 3.00 & B4V \\
 260131665 & HD 42933          & EB (EB/D:)      &  4.810  &       &       &     &   &      &  170 &  25981 &  6 & 4.58 & B0.5:III?np + B0.5/3:\\
 260368525 & HD 44937          & ROT?            &  8.190  & 1.194 &    31 &  16 & 1 &  241 &      &  10500 &  5 & 2.24 & B9.5V \\
 260540898 & HD 46212          & ROT?            &  8.260  & 0.562 &    30 &  11 & 1 &  117 &      &  12400 &  5 & 2.56 & B8IV \\
 260640910 & HD 46860          & BE              &  5.707  &       &       &     &   &      &  200 &  13520 &  1 & 2.70 & B8III (Be) \\
 260820871 & HD 218801         & EP              &  8.990  &       &       &     &   &      &      &  10500 &  5 & 1.92 & B9.5Vn: \\
 261205462 & HD 40953          & SPB             &  5.451  &       &       &     &   &      &   23 &  11243 &  4 & 1.93 & B9V \\
 262815962 & HD 218976         & ROT             &  8.120  & 0.369 &   207 &  58 & 1 &   40 &      &   9846 &  6 & 1.60 & B9.5/A0V \\
 270070443 & HD 198174         & SXARI           &  5.854  & 0.395 &  2529 &  99 & 0 &   64 &   72 &  13217 & 13 & 2.46 & B7IIIp  \\
 270219259 & HD 209014         & BE+MAIA         &  5.620  &       &       &     &   &      &  350 &  13520 &  1 & 2.96 & B8III shell (Be) \\
 270557257 & HD 49835          & ROT?            &  8.560  & 2.475 &    42 &  13 & 0 &  253 &      &  10500 &  5 & 1.65 & B9.5V \\
 270622440 & HD 224112         & -               &  6.828  &       &       &     &   &      &   35 &        &    & 2.32 & Blend with 270622446 \\
 270622446 & HD 224113         & EA (EA/DM)      &  6.087  &       &       &     &   &      &  135 &  13665 &  4 & 2.70 & B7(V) + B9(V) \\
 271503441 & HD 2884           & ROT             &  4.335  & 3.056 &   234 &  51 & 1 &  282 &  140 &  11576 & 14 & 1.73 & B8/A0 \\
 271971626 & HD 62153          & ROT+MAIA        &  7.020  & 0.215 &   289 &  88 & 0 &   38 &      &  11783 &  4 & 2.33 & B9IV \\
\hline
\end{tabular}
}
\end{center}
\end{table*}

\begin{table*}
\begin{center}
\resizebox{18cm}{!}{
\begin{tabular}{rllrrrrrrrrrrl}
\hline
\multicolumn{1}{c}{TIC}               & 
\multicolumn{1}{l}{Name}              & 
\multicolumn{1}{l}{Var. Type}         & 
\multicolumn{1}{c}{$V$}               & 
\multicolumn{1}{c}{$\nu_{\rm rot}$}   & 
\multicolumn{1}{c}{$A_{\rm rot}$}     & 
\multicolumn{1}{c}{S/N}     & 
\multicolumn{1}{c}{$N_H$}             & 
\multicolumn{1}{c}{$v_e$}             & 
\multicolumn{1}{c}{$v\sin i$}         & 
\multicolumn{1}{c}{$T_{\rm eff}$}     & 
\multicolumn{1}{c}{Ref}               & 
\multicolumn{1}{c}{$\log\tfrac{L}{L_\odot}$} & 
\multicolumn{1}{l}{Sp. Type}          \\
\multicolumn{1}{c}{}                  & 
\multicolumn{1}{l}{}                  & 
\multicolumn{1}{l}{}                  & 
\multicolumn{1}{c}{mag}               &
\multicolumn{1}{c}{(d$^{-1}$)}        & 
\multicolumn{1}{c}{(ppm)}             & 
\multicolumn{1}{c}{}                  & 
\multicolumn{1}{c}{}                  & 
\multicolumn{1}{c}{(km/s)}            & 
\multicolumn{1}{c}{(km/s)}            & 
\multicolumn{1}{c}{(K)}               &
\multicolumn{1}{c}{}                  &
\multicolumn{1}{c}{}                  &
\multicolumn{1}{l}{}                 \\
\hline
 276864600 & HD 269777         & ACYG            & 11.060  &       &       &     &   &      &      &  21370 &  1 &      & B3Ia (LMC) \\
 277022505 & HD 269786         & ACYG            & 11.180  &       &       &     &   &      &      &  28000 &  1 &      & B1I (LMC) \\
 277022967 & HD 37836          & ACYG            & 10.660  &       &       &     &   &      &      &  28000 &  1 &      & B0e(q)I (LMC) \\
 277099925 & HD 269845         & ACYG            & 11.790  &       &       &     &   &      &      &  22580 &  1 &      & B2.5Ia (LMC) \\
 277103567 & HD 37935          & ROT (BE)        &  6.281  & 1.496 &   209 &  74 & 3 &  361 &  209 &   9940 &  6 & 2.30 & B9.5V (Be)\\
 277172980 & HD 37974          & ACYG            & 10.959  &       &       &     &   &      &      &  28000 &  1 &      & B0.5e (LMC) \\
 277173650 & HD 269859         & ACYG            & 10.730  &       &       &     &   &      &      &  23790 &  1 &      & B1.5Ia  (LMC) \\
 277298891 & Sk-69 237         & ACYG            & 12.080  &       &       &     &   &      &      &  24625 &  2 &      & B1Ia (LMC) \\
 277982164 & HD 54239          & ROT             &  5.459  & 1.266 &   122 &  66 & 1 &  248 &  209 &   9938 &  4 & 2.12 & B9.5/A0III/IV \\
 278683664 & HD 47770          & ROT?            &  8.490  & 1.701 &    32 &  13 & 0 &  268 &      &   9012 &  7 & 1.76 & B9.5V \\
 278865766 & HD 48971          & -               &  8.280  &       &       &     &   &      &      &   9743 &  7 & 1.93 & B9V \\
 278867172 & HD 49111          & -               &  8.490  &       &       &     &   &      &      &  14273 &  7 & 2.11 & B9.5V \\
 279430029 & HD 53048          & ROT             &  7.920  & 1.784 &    64 &  33 & 1 &  554 &      &  18950 &  1 & 3.64 & B5/7Vn(e:) (Be) \\
 279511712 & HD 53921          & ELL (LPB)       &  5.600  & 0.605 &  4590 & 141 & 1 &  141 &      &  13800 &  4 & 2.84 & B9III+B8V \\
 279957111 & HD 269582         & -               & 12.597  &       &       &     &   &      &      &  30200 &  1 &      & Ofpe/WN9 (LMC) \\
 280051467 & HD 19400          & SXARI           &  5.497  & 0.229 &   570 & 135 & 0 &   36 &   64 &  14117 &  6 & 2.56 & B8III/IV Hewk \\
 280684074 & HD 215573         & SPB+ROT (LPB)   &  5.313  & 0.563 & 10068 &  77 & 0 &  104 &   13 &  13960 & 15 & 2.66 & B6V \\
 281703963 & HD 4150           & MAIA            &  4.365  &       &       &     &   &      &  105 &   9822 & 14 & 2.06 & A0IV/B9Vp((SiFe)) \\
 281741629 & CD-56 152         & BE (BE)         & 10.180  &       &       &     &   &      &  180 &  19000 & 16 & 4.59 & sdB?/Be? \\
 293268667 & HD 47478          & SPB+ROT         &  8.500  & 3.390 &   386 &  75 & 1 &  455 &      &  10765 &  7 & 1.93 & B9V \\
 293271581 & Hen 3-15          & EB (NB)         & 12.502  &       &       &     &   &      &      &        &    & 0.67 & Bem RR Pic (Nova) \\
 293973218 & HD 54967          & SPB             &  6.470  &       &       &     &   &      &   34 &  22570 &  1 & 3.47 & B3III \\
 294747615 & HD 30612          & SXARI           &  5.515  & 0.192 &   670 & 130 & 0 &   30 &   30 &  13661 &  6 & 2.50 & B8II/IIIp:Si: \\
 294872353 & HD 270754         & ACYG            & 11.260  &       &       &     &   &      &      &  19910 &  2 &      & B1.5Ia: (LMC) \\
 300010961 & HD 55478          & ROT+MAIA        &  8.060  & 0.683 &   723 & 117 & 3 &  122 &      &  11144 &  7 & 2.24 & B8III \\
 300325379 & HD 58916          & ROT             &  8.010  & 1.838 &    42 &  20 & 0 &  268 &      &  11710 &  1 & 2.15 & B9Vn \\
 300329728 & HD 59426          & ELL/ROT         &  8.420  & 0.411 &  5216 & 140 & 1 &   51 &      &  11710 &  1 & 2.02 & B9V \\
 300744369 & HD 63928          & ROT             &  8.700  & 0.957 &   475 & 115 & 1 &  102 &      &  11710 &  1 & 1.88 & B9V \\
 300865934 & HD 64484          & -               &  5.774  &       &       &     &   &      &      &  10707 &  4 & 2.14 & B9V \\
 306672432 & HD 67252          & ROT?            &  8.530  & 1.090 &    31 &  13 & 1 &  119 &      &  13520 &  1 & 2.15 & B8/9V \\
 306824672 & HD 68221          & SPB?            &  8.650  &       &       &     &   &      &      &  11710 &  5 & 2.32 & B9V \\
 306829961 & HD 68520          & SPB             &  4.400  &       &       &     &   &      &   10 &  14090 & 17 & 3.44 & B5III \\
 307291308 & HD 71066          & SXARI           &  5.617  & 0.387 &   300 &  42 & 1 &   62 &    2 &  11821 &  6 & 2.25 & B9pMnHg \\
 307291318 & HD 71046          & -               &  5.329  &       &       &     &   &      &   69 &  12102 &  6 & 2.37 & B9III/IV \\
 307993483 & HD 73990          & SPB+ROT?        &  6.870  & 3.697 &    48 &  27 & 0 &      &      &  10221 &  7 & 2.02 & B7/8V \\
 308395911 & HD 66591          & SPB             &  4.797  &       &       &     &   &      &   43 &  16983 &  6 & 2.96 & B3IV \\
 308454245 & HD 67420          & MAIA            &  8.250  &       &       &     &   &      &      &  11710 &  1 & 2.15 & B9V \\
 308456810 & HD 67170          & ROT             &  8.110  & 0.251 &    54 &  23 & 1 &   38 &      &  13520 &  1 & 2.43 & B8III/IV \\
 308537791 & HD 67277          & ROT+MAIA        &  8.260  & 0.527 &   746 &  81 & 1 &   78 &      &  13520 &  1 & 2.41 & B8III \\
 308748912 & HD 68423          & -               &  6.313  &       &       &     &   &      &      &  15330 &  1 & 2.86 & B7IVek (Be)\\
 309702035 & HD 271163         & ACYG            &  9.984  &       &       &     &   &      &      &  21370 &  1 &      & B3Ia (LMC)\\
 313934087 & HD 224990         & SPB             &  5.023  &       &       &     &   &      &   15 &  16100 & 17 & 3.08 & B4III \\
 327856894 & HD 225253         & -               &  5.581  &       &       &     &   &      &      &  13123 &  6 & 2.64 & B8IV/V \\
 349829477 & HD 61267          & -               &  8.330  &       &       &     &   &      &      &  10000 &  1 & 1.72 & B9/A0IV \\
 349907707 & HD 61644          & EA (EA)         &  8.410  &       &       &     &   &      &      &  18000 &  1 & 2.76 & B5/6IV \\
 350146577 & HD 63204          & SXARI           &  8.310  & 0.544 & 48570 & 139 & 1 &   78 &      &  10505 & 10 & 1.95 & B9Si \\
 350823719 & HD 41037          & EB              &  9.460  &       &       &     &   &      &  215 &  22570 &  1 & 3.70 & B3V \\
 354671857 & HD 14228          & ROT?            &  3.570  & 2.908 &   256 &  98 & 0 &  375 &  200 &  12687 &  6 & 2.18 & B8IV-V \\
 355141264 & HD 208495         & -               &  8.860  &       &       &     &   &      &      &   9945 &  7 & 1.58 & B9.5V \\
 355477670 & HD 220802         & ROT?            &  6.210  & 0.272 &    21 &  23 & 0 &   37 &  123 &  11206 &  4 & 2.03 & B9V \\
 355653322 & HD 224686         & ROT             &  4.470  & 1.266 &   117 &  38 & 1 &  274 &  275 &  11710 &  1 & 2.49 & B9IIIn (Be) \\
 358232450 & HD 6882           & EA (EA/DM)      &  3.967  &       &       &     &   &      &  111 &  13471 &  6 & 2.41 & B6V + B9V\\
 358466708 & CD-60 1931        & ROT             &  8.090  & 1.661 &  3405 &  83 & 1 &  280 &  223 &  12814 & 18 & 2.43 & B8III \\
 358467049 & CPD-60 944        & SXARI           &  8.756  & 0.265 & 13161 &  97 & 0 &   30 &      &  12600 & 18 & 2.07 & B8pSi\\
 358467087 & CD-60 1929        & SPB?            &  8.520  &       &       &     &   &      &   43 &  12543 & 18 & 2.26 & B8.5IV \\
 364323837 & HD 40031          & SPB?            &  9.270  &       &       &     &   &      &   65 &  14000 & 16 & 3.08 & sdB, B6III \\
 364398190 & CD-60 1978        & ROT             &  8.750  & 0.760 &   129 &  28 & 0 &  101 &   68 &  12337 & 18 & 2.16 & B8.5IV-V \\
 364398342 & HD 66194          & BE (GCAS)       &  5.810  &       &       &     &   &      &  200 &  20632 & 18 & 3.76 & B2IVn(e)p(Si) (Be) \\
 364421326 & HD 66109          & ROT?            &  8.190  & 1.976 &    31 &  11 & 0 &  385 &      &  11710 &  1 & 2.40 & B9.5V \\
 369397090 & CD-30 19716       & -               & 12.860  &       &       &     &   &      &      &  39811 & 19 & 1.43 & sdB \\
 369457005 & HD 197630         & SPB?            &  5.474  &       &       &     &   &      &      &  11511 &  6 & 1.99 & B8/9V \\
 370038084 & HD 26109          & -               &  8.580  &       &       &     &   &      &      &  11710 &  1 & 1.62 & B9.5/A0V \\
 372913233 & HD 65950          & -               &  6.870  &       &       &     &   &      &   26 &  12842 & 18 & 2.88 & B8.5IIIpMnHg \\
 372913582 & CD-60 1954        & SPB?            &  8.590  &       &       &     &   &      &  188 &  10579 & 18 & 2.13 & B9.5V \\
 372913684 & HD 65987          & SXARI (EA:)     &  7.590  & 0.685 &  5892 &  93 & 2 &  163 &   13 &  12600 & 18 & 2.70 & B9.5IVpSi \\
 373843852 & HD 269525         & ACYG            & 12.780  &       &       &     &   &      &      &  28000 &  1 &      & B0: (LMC) \\
 389921913 & HD 270196         & ACYG            & 11.600  &       &       &     &   &      &      &  22875 &  2 &      & B1.5Ia (LMC) \\
 391810734 & HD 269655         & ACYG            & 12.200  &       &       &     &   &      &      &  28800 & 20 &      & B0Ia (LMC) \\
 391887875 & HD 269660         & ACYG            & 11.190  &       &       &     &   &      &      &  23790 &  1 &      & B1.5Ia (LMC) \\
 404768847 & VFTS 533          & ACYG?           & 11.820  &       &       &     &   &      &   57 &  19275 &  2 &      & B1.5Ia+qp (LMC)\\
 404768956 & NGC2070 Mel 12    & ACYG?           & 11.996  &       &       &     &   &      &      &  28510 & 20 &      & B0/0.5Ia (LMC) \\
 404796860 & HD 269920         & ACYG            & 11.650  &       &       &     &   &      &      &  18950 &  1 &      & B5Ia (LMC) \\
 404852071 & Sk-69 265         & ACYG            & 11.880  &       &       &     &   &      &      &  21370 &  1 &      & B3Ia (LMC) \\
 404933493 & HD 269997         & ACYG            & 11.200  &       &       &     &   &      &      &  22580 &  1 &      & B2.5Ia (LMC) \\
 404967301 & HD 269992         & ACYG            & 11.220  &       &       &     &   &      &      &  18020 &  2 &      & B2.5:Ia: (LMC) \\
 410447919 & HD 64811          & ROT?            &  8.450  & 0.219 &   317 &  58 & 0 &  133 &      &  20760 &  1 & 4.38 & B4III \\
 410451677 & HD 66409          & SXARI           &  8.410  & 0.488 &  1292 &  98 & 0 &   67 &   34 &  12987 & 18 & 2.28 & B8.5IV(HgMn)? \\
 419065817 & HD 1256           & ROT             &  6.488  & 1.474 &   138 &  52 & 3 &  186 &  166 &  14280 &  6 & 2.37 & B6III/IV \\
 425057879 & HD 269676         & EB/ELL?         & 11.550  &       &       &     &   &      &  120 &  41000 &  1 &      & O6+O9 (LMC) \\
\hline
\end{tabular}
}
\end{center}
\end{table*}

\begin{table*}
\begin{center}
\resizebox{18cm}{!}{
\begin{tabular}{rllrrrrrrrrrrl}
\hline
\multicolumn{1}{c}{TIC}               & 
\multicolumn{1}{l}{Name}              & 
\multicolumn{1}{l}{Var. Type}         & 
\multicolumn{1}{c}{$V$}               & 
\multicolumn{1}{c}{$\nu_{\rm rot}$}   & 
\multicolumn{1}{c}{$A_{\rm rot}$}     & 
\multicolumn{1}{c}{S/N}     & 
\multicolumn{1}{c}{$N_H$}             & 
\multicolumn{1}{c}{$v_e$}             & 
\multicolumn{1}{c}{$v\sin i$}         & 
\multicolumn{1}{c}{$T_{\rm eff}$}     & 
\multicolumn{1}{c}{Ref}               & 
\multicolumn{1}{c}{$\log\tfrac{L}{L_\odot}$} & 
\multicolumn{1}{l}{Sp. Type}          \\
\multicolumn{1}{c}{}                  & 
\multicolumn{1}{l}{}                  & 
\multicolumn{1}{l}{}                  & 
\multicolumn{1}{c}{mag}               &
\multicolumn{1}{c}{(d$^{-1}$)}        & 
\multicolumn{1}{c}{(ppm)}             & 
\multicolumn{1}{c}{}                  & 
\multicolumn{1}{c}{}                  & 
\multicolumn{1}{c}{(km/s)}            & 
\multicolumn{1}{c}{(km/s)}            & 
\multicolumn{1}{c}{(K)}               &
\multicolumn{1}{c}{}                  &
\multicolumn{1}{c}{}                  &
\multicolumn{1}{l}{}                 \\
\hline
 425064757 & HD 269696         & EA (EA/D)       & 11.138  &       &       &     &   &      &      &  42000 & 19 &      & sdO (LMC) \\
 425081475 & HD 269700         & ACYG            & 10.540  &       &       &     &   &      &      &  23790 &  1 &      & B1.5Iaeq (LMC) \\
 425083410 & HD 269698         & ACYG            & 12.220  &       &       &     &   &      &      &  40400 &  1 &      & O4If (LMC) \\
 425084841 & TYC 8891-3638-1   & ACYG            & 12.180  &       &       &     &   &      &  62  &  23660 &  2 &      & B1Ia (LMC) \\
 441182258 & HD 210934         & SPB+ROT?        &  5.430  & 1.274 &    66 &  26 & 0 &  235 &   56 &  12526 &  6 & 2.47 & B8III \\
 441196602 & HD 211993         & ROT?            &  8.200  & 0.144 &    49 &  15 & 0 &   20 &      &  10750 &  6 & 1.96 & B8/9V \\
 469906369 & HD 212581         & MAIA            &  4.495  &       &       &     &   &      &  200 &  11271 &  8 & 2.18 & B9.5IVn \\
\hline
\end{tabular}
}
\end{center}
\end{table*}

\begin{table}
\begin{center}
\caption{References to the effective temperatures in Table\,\ref{bstars}.}
\label{refs}
\begin{tabular}{rll}
\hline
\multicolumn{1}{c}{Ref}               & 
\multicolumn{1}{c}{Reference}         & 
\multicolumn{1}{c}{Method}            \\
\hline
    1 & \citet{Pecaut2013}         & Spectral type        \\   
    2 & \citet{Urbaneja2017}       & Spectroscopy        \\
    3 & \citet{Gianninas2011}      & Spectroscopy        \\
    4 & \citet{Paunzen2005}        & Narrow-band photometry \\
    5 & \citet{Wright2003}         & Spectral type        \\
    6 & \citet{Balona1994a}        & Str\"{o}mgren        \\
    7 & \citet{Chandler2016}       & SED fitting          \\
    8 & \citet{Gullikson2016}      & Spectroscopy        \\
    9 & \citet{SanchezBlazquez2006}& Spectroscopy        \\
   10 & \citet{McDonald2017}       & SED fitting          \\
   11 & \citet{Geier2017}          & Stectral type/SED    \\
   12 & \citet{Oey1995}            & Sp.Type and UBV      \\
   13 & \citet{Paunzen2013}        & UBV/Geneva/Str\"{o}mgren \\
   14 & \citet{David2015}          & Str\"{o}mgren        \\
   15 & \citet{DeCat2002}          & Geneva               \\
   16 & \citet{Silva2011}          & Str\"{o}mgren        \\
   17 & \citet{Zorec2009}          & BCD method           \\
   18 & \citet{Silaj2014}          & Geneva/Str\"{o}mgren \\
   19 & \citet{Soubiran2016}       & Spectral type        \\
   20 & \citet{Massey2002}         & UBVR photometry      \\
   21 & \citet{Niemczura2015}      & Spectroscopy         \\
   22 & \citet{Huber2016}          & Spectroscopy        \\
   23 & \citet{Balona2011b}        & Spectroscopy        \\
   24 & \citet{Tkachenko2013b}     & Spectroscopy        \\
\hline
\end{tabular}
\end{center}
\end{table}

\section{Other types of variability in B stars}

There are two main types of pulsating variable among  the B stars, the 
$\beta$\,Cep (called BCEP in the GCVS) and ``slowly pulsating B-star" (SPB) 
variables.   The pulsations in both types are caused by the opacity mechanism 
in the ionization zone of iron-group elements \citep{Dziembowski1993b}.  The 
$\beta$~Cep stars are hotter (20000--32000\,K) and pulsate with frequencies in
the range 4--12\,d$^{-1}$, while the SPB stars are cooler (11000-19000\,K) and 
pulsate with frequencies in the range 0.3--3\,d$^{-1}$. 

In the GCVS, SPB stars are given the variability type LPB (``long-period 
B-star'').  The reason for the different designation is that the 
GCVS wisely tries to avoid naming a variable type according to a specific 
interpretation of the cause of the light variation.  If a much better 
interpretation for SPB light variations were to be found, for example, then 
the class will have to be renamed.  However, since the designation LPB has
fallen into disuse, SPB is used instead.

The ACYG variables are nonradially pulsating B-type supergiants.  The
multiple periods range from several days to several weeks.  The B supergiants 
observed by {\it TESS} are all members of the Large Magellanic Cloud (LMC).  
With few exceptions, the light curves of all the LMC supergiants matched those
expected for the ACYG variables.

The Be variables are B stars which show, or have shown, emission in H$\alpha$ 
or other Balmer lines.  This can include many different types of object,
including supergiants, in which the emission is thought to be due to
different physical mechanisms.  The ``classical Be stars'' are a narrower
set confined to stars in the main sequence band.  Some Be stars show large, 
frequent outbursts in the light curve attributed to the sudden ejection of 
circumstellar material (GCAS variables).  Others are less active and show 
quasi-periodic variations with timescales in the range 0.5--2\,d, usually 
attributed to multiple nonradial pulsations and/or obscuration by 
circumstellar material.  In the GCVS, the designation BE is used to describe 
these stars.  The photometric variability type BE should not be confused with 
the spectroscopic classification ``Be'' which designates emission in some
Balmer lines.

The EA and EB variables are eclipsing binaries.  In EA variables it is 
possible to specify the beginning and end of eclipses.  Between eclipses the 
light is more-or-less constant.  On the other hand, the EB eclipsing variables 
are close binaries where the variation is practically sinusoidal with no
constant light.  The EP stars show very small eclipses which may be 
attributed to a planet or sub-stellar companion.

The MAIA type is a new class which was introduced by \citet{Balona2015c,
Balona2016c}.  These are B stars which show many high-frequency peaks
similar to those seen in BCEP or $\delta$~Scuti variables (called DSCT in
the GCVS).  However, they are too cool to be classified as BCEP and too hot to
be DSCT. Whether or not they deserve a separate class remains to be seen. 
These stars may be related to the ``FaRPB'' stars \citep{Mowlavi2016} which 
also show high frequencies and lie between the $\beta$~Cep and $\delta$~Scuti 
variables.  The FaRPB stars, however, are all rapidly-rotating stars. The 
evidence suggests that MAIA stars are not rapid rotating stars 
\citep{Balona2016c}.

Some subdwarf B stars (sdB stars) are also known to be multiperiodic variables.
There are two classes: the V361\,Hya and the V1093\,Her stars.  V361\,Hya stars 
are short-period pulsating sdB stars with $24000 < T_{\rm eff} < 40000$\,K and 
$\log g < 5.8$ (less compact than white dwarfs).  They have multiple periods 
in the range 60--400 \,s (200--1500\,d$^{-1}$). They are also known as 
EC\,14026 stars.   The V1093\,Her stars are in the same general area of the 
H-R diagram, but somewhat cooler and less compact. They are long-period
(1800--9000\,s or 10--50\,d$^{-1}$) analogues of the V361\,Hya stars and are 
also known as PG\,1716 stars.

\begin{figure}
\centering
\includegraphics[]{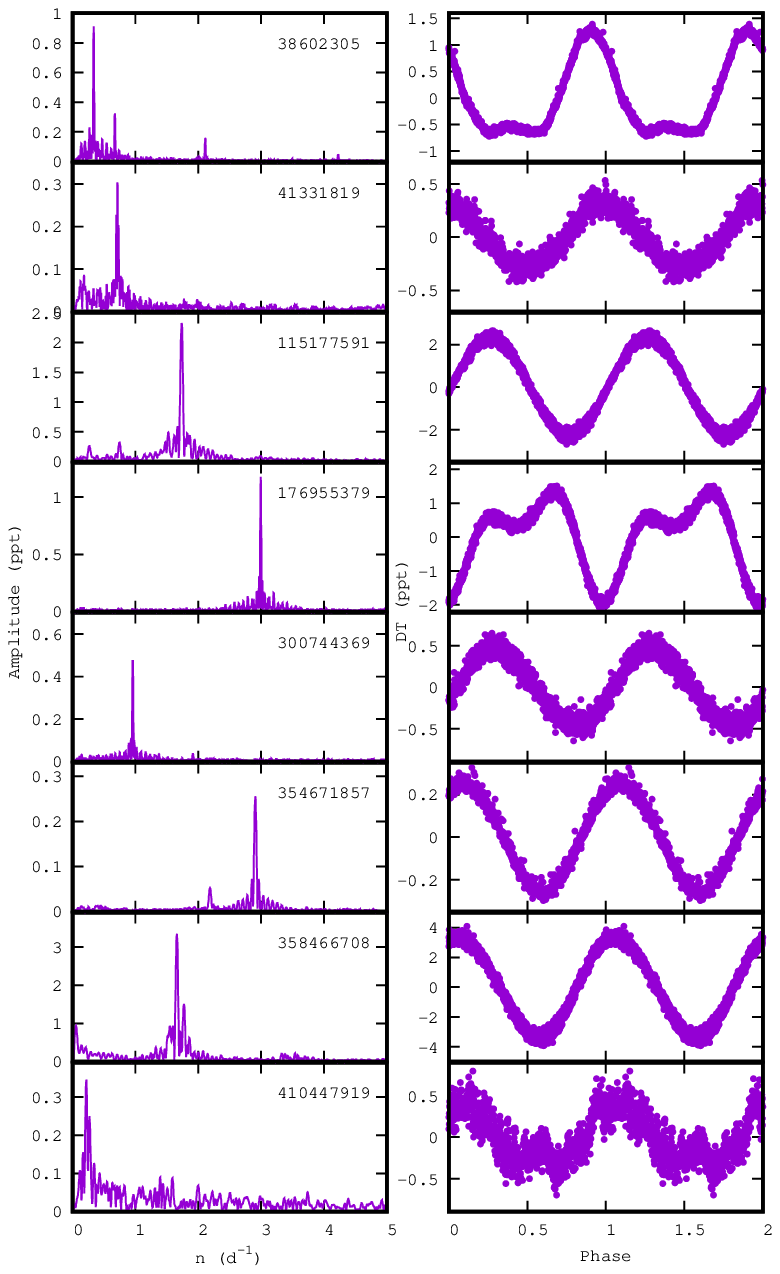}
\caption{Examples of the periodograms (left panels) and phased {\it TESS}
light curves (right panels) of a presumed rotational variables.}
\label{phase}
\end{figure}

Table\,\ref{bstars} lists our assigned variability type in the third column.
Where the star seems constant or no definite assignment is possible, a dash
is used.  When the classification is uncertain a question mark is added
(e.g. ROT?). Sometimes two classifications are possible and either is
acceptable.  This is shown by a slash, e.g: ROT/ELL.  In other cases two
types seem to be present in the same star which is shown by a plus, e.g. 
SPB+ROT.  The classification of ROT in a pulsating star such as SPB or BCEP
is sometimes made if a strong peak and its harmonic are present.  The
justification is that there is no reason why rotational modulation cannot
co-exist with pulsation.   The harmonic may, of course, also arise as a
result of nonlinear pulsation.  Fig.{\,\ref{phase}} shows some examples of the 
periodograms and the phased light curves of stars classified as a ROT 
variables.

\section{Effective temperature}

The effective temperature, $T_{\rm eff}$, may be estimated in several 
different ways, sometimes giving very different results.  Literature values
of $T_{\rm eff}$ were chosen according to the following order of descending
priority.  

Whenever possible, estimates of $T_{\rm eff}$ from modelling the stellar
spectrum were chosen.  Failing this, an  estimate based on narrow-band 
photometry (i.e.Str\"{o}mgren $uvby\beta$ or Geneva photometry) was used 
For some stars, Str\"{o}mgren photometry was available, but no value of 
$T_{\rm eff}$ had yet been derived.  In such cases we estimated $T_{\rm eff}$ 
by de-reddening the star using the method described by \citet{Crawford1978} and
then applying the calibrations of \citet{Balona1994a} to obtain $T_{\rm eff}$.  

Next in priority were methods using the spectral energy distribution (SED)
from wide-band photometry, usually UBVRI. There can be problems with this 
method if measurements in the U band are missing, so $T_{\rm eff}$ from SED was
selected only if it agreed reasonably well with the temperature from the 
spectral type.  If not, or if no $T_{\rm eff}$ measurements could be found, the
spectral type itself was used to estimate $T_{\rm eff}$ from the table of 
\citet{Pecaut2013}. For emission-line stars the $T_{\rm eff}$ from the 
spectral type  was used instead of photometric methods.  

The error in $T_{\rm eff}$ clearly depends on the method used, but we can 
obtain an approximate overall value for B stars from the PASTEL catalogue 
\citep{Soubiran2016}.  The errors increase with $T_{\rm eff}$ ranging from 
500 to 1500\,K.  We adopt a standard error of 1000\,K as reasonable overall 
estimate.

\section{Luminosities and radii}

From the {\it Gaia} DR2 parallax $\pi$ \citep{Gaia2016, Gaia2018}, the 
absolute magnitude is calculated using 
\mbox{$M_V = V_0 + 5(\log_{10}\pi + 1)$}, where $V_0$ is the 
reddening-free $V$ magnitude. We used $V$ magnitudes from SIMBAD. The 
reddening correction was derived from a three-dimensional reddening map with a 
radius of 1200\,pc around the Sun and within 600\,pc of the galactic midplane 
as calculated by \citet{Gontcharov2017}.  For more distant stars, a simple 
reddening model is used (see Eq. 20 of \citealt{Brown2011a}), but adjusted so 
that it agrees with the 3D map at 1200\,pc.  

The absolute bolometric magnitude is given by $M_{\rm bol} = M_V + 
{\rm BC}_V - M_{\rm bol\odot}$, where ${\rm BC_V}$ is the bolometric
correction in $V$ and  $M_{\rm bol\odot} = 4.74$ is the absolute bolometric
magnitude of the Sun. The bolometric correction as a function of
$T_{\rm eff}$ is given in \citet{Pecaut2013}.   Finally, the 
luminosity relative to the Sun is found using  $\log L/L_\odot =  
-0.4M_{\rm bol}$.  {\it Gaia} DR2 parallaxes for many early-type stars are 
subject to larger errors than quoted because the match with the astrometric
 model used to determine the parallax is rather poor, perhaps due to binarity 
\citep{Gaia2018}.  Nevertheless, these are the best parallax estimates at 
present.  From the error in the {\it Gaia} DR2 parallax, the typical standard 
deviation in $\log(L/L_\odot)$ is estimated to be about 0.05\,dex, allowing for
standard deviations of 0.01\,mag in the apparent magnitude, 0.10\,mag in visual
extinction and 0.02\,mag in the bolometric correction in addition to the 
parallax error.  The true error in luminosity is likely larger for the reason 
just quoted.

From the luminosity and effective temperature, the stellar radius, 
$R/R_\odot$, can be found.  For stars where the rotational modulation 
frequency $\nu_{\rm rot}$ is available, the equatorial rotational velocity 
$v_e$ can be determined.  Table\,\ref{bstars} shows $\log L/L_\odot$.  For 
those stars with known $\nu_{\rm rot}$, $v_e$ is also shown.

\begin{figure}
\centering
\includegraphics[]{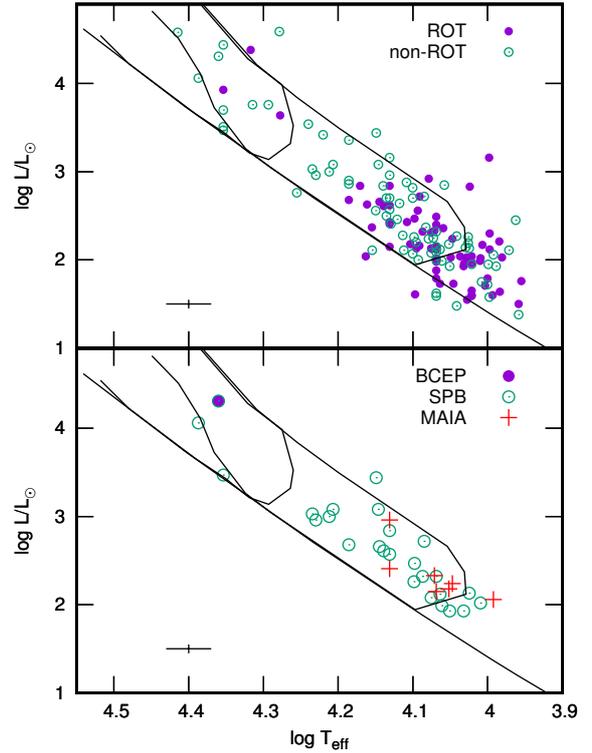}
\caption{The theoretical H-R diagram for {\it TESS, Kepler} and {\it K2} 
main sequence stars.  The top panel shows stars identified as 
possible rotational variables (filled circles).  The open circles are other 
main sequence stars (including the SXARI rotational variables).  In the
bottom panel pulsating variable stars in the {\it TESS} field are shown.  The
filled circle is the only BCEP star in the sample, HN~Aqr.  The open circles 
are stars classified as SPB variables.  The crosses are MAIA stars.  The 
solid line is the zero-age main sequence from solar-abundance models by 
\citet{Bertelli2008}.  The dashed and dotted areas are the theoretical BCEP
and SPB instability regions from \citet{Miglio2007}.  The cross at the
bottom shows the typical 1-$\sigma$ error bars.} 
\label{hrdiag}
\end{figure}

In addition to the {\it TESS} stars, we have examined the light curves of {\it
Kepler} and {\it K2} data for possible rotational modulation.   The {\it
Kepler} data have a time span of nearly 4\,yr which results in a very low
periodogram noise level.  The {\it K2} data have a timespan of around 80\,d.  
Table\,\ref{add} lists the measurements and stellar parameters obtained in the 
same way.  Fig.\,\ref{hrdiag} shows the main sequence stars in 
Tables\,\ref{bstars} and \ref{add} in the theoretical H-R diagram.

\begin{table*}
\begin{center}
\caption{Additional rotational variables identified from the light curves of
the {\it Kepler} and {\it K2} missions.}
\label{add}
\resizebox{18cm}{!}{
\begin{tabular}{lllrrrrrrrrrrl}
\hline
\multicolumn{1}{l}{KIC/EPIC}          & 
\multicolumn{1}{l}{Name}              & 
\multicolumn{1}{l}{Var. Type}         & 
\multicolumn{1}{c}{$V$}               & 
\multicolumn{1}{c}{$\nu_{\rm rot}$}   & 
\multicolumn{1}{c}{$A_{\rm rot}$}     & 
\multicolumn{1}{c}{S/N}             & 
\multicolumn{1}{c}{$N_H$}             & 
\multicolumn{1}{c}{$v_e$}             & 
\multicolumn{1}{c}{$v\sin i$}         & 
\multicolumn{1}{c}{$T_{\rm eff}$}     & 
\multicolumn{1}{c}{Ref}               &
\multicolumn{1}{c}{$\log\tfrac{L}{L_\odot}$} & 
\multicolumn{1}{l}{Sp. Type}             \\
\multicolumn{1}{l}{}                  & 
\multicolumn{1}{l}{}                  & 
\multicolumn{1}{l}{}                  & 
\multicolumn{1}{c}{mag}               &
\multicolumn{1}{c}{(d$^{-1}$)}        & 
\multicolumn{1}{c}{(ppm)}             & 
\multicolumn{1}{c}{}                  & 
\multicolumn{1}{c}{}                  & 
\multicolumn{1}{c}{(km/s)}            & 
\multicolumn{1}{c}{(km/s)}            & 
\multicolumn{1}{c}{(K)}               &
\multicolumn{1}{c}{}                  &
\multicolumn{1}{c}{}                  &
\multicolumn{1}{l}{}                  \\
\hline
EPIC 210788932  & HD 23016       &  ROT          &  5.690  & 1.802 &  2743 & 39 &  0 &  351 & 260  & 11463 &  6 & 2.36 & B8Ve? (Be) \\
EPIC 211116936  & HD 23324       &  SPB+ROT      &  5.640  & 1.543 &   161 & 19 &  1 &  252 & 206  & 12218 &  6 & 2.32 & B8:IV/V \\
EPIC 211028385  & HD 23753       &  ROT          &  5.450  & 1.808 &    57 & 18 &  1 &  308 & 292  & 11899 &  6 & 2.31 & B8IV/V  \\
EPIC 211054599  & HD 23964       &  SXARI (UV)   &  6.740  & 0.633 &   470 & 26 &  1 &   77 &  21  & 10190 &  6 & 1.75 & B9.5VspSiSrCr \\
EPIC 210964459  & HD 26571       &  SXARI (ACV:) &  6.120  & 0.063 &  8910 & 40 &  0 &   22 &  22  & 11430 &  6 & 2.85 & B9IIIp:(Si) \\
EPIC 202061205  & HD 253049      &  ROT          &  9.620  & 0.294 &  1714 & 23 &  1 &   90 & 184  & 22570 &  1 & 3.93 & B3III \\
EPIC 211311439  & HD 74521       &  SXARI (ACV)  &  5.660  & 0.144 &  3938 & 10 &  0 &   29 &  18  & 10615 &  6 & 2.26 & B9pSiCr \\
EPIC 201232619  & HD 97991       &  ROT          &  7.410  & 0.496 &   128 & 17 &  1 &  323 & 137  &  9956 &  6 & 3.16 & B2/3V \\
EPIC 204760247  & HD 142883      &  ROT          &  5.841  & 1.103 &  2562 & 18 &  0 &  254 &  19  &  9648 &  6 & 2.21 & B3V \\
EPIC 204134887  & HD 142884      &  SXARI (ACV)  &  6.777  & 1.245 &  5815 & 35 &  2 &  200 & 138  &  9979 &  6 & 1.95 & kB8hB4HeB9V Si4200\\
EPIC 204348206  & HD 143600      &  ROT (RS:)    &  7.330  & 3.922 &   205 & 33 &  3 &  452 & 265  &  9087 &  6 & 1.50 & B9Vann \\
EPIC 204095429  & HD 144844      &  SXARI        &  5.880  & 0.116 &  1002 & 36 &  1 &   25 &  98  &  9362 &  6 & 2.11 & B9MnPGa\\
EPIC 204964091  & HD 147010      &  SXARI (ACV)  &  7.400  & 0.255 & 12814 & 42 &  1 &   25 &  25  &  9092 &  6 & 1.38 & B9SiCrSr*\\
EPIC 205417334  & HD 148860      &  ROT          &  8.040  & 0.920 &   152 & 21 &  1 &  177 & 257  &  9566 &  6 & 2.03 & B9.5V \\
KIC 8351193     & HD 177152      &  ROT          &  7.570  & 1.757 &     2 & 10 &  1 &  168 & 180  & 10500 & 21 & 1.59 & A0VkB8mB7 $\lambda$ Boo\\
EPIC 217692814  & HD 177015      &  ROT          &  7.800  & 1.139 &    50 & 13 &  1 &  448 & 202  & 10567 & 22 & 2.83 & B5V(e) (Be) \\
KIC 8087269     & ILF1 +43 30    &  ROT          & 11.710  & 1.610 &  1608 & 93 &  3 &  268 & 271  & 14500 & 23 & 2.63 & B5 \\
KIC 9278405     & ILF1 +45 284   &  SPB/ROT      & 10.160  & 1.805 &     4 & 10 &  1 &  194 & 110  & 11710 &  1 & 1.79 & B9 \\
KIC 4056136     & BD+38 3580     &  ROT          &  9.550  & 2.370 &    10 & 39 &  1 &  351 & 227  & 10500 & 21 & 1.97 & B9IV-Vnn \\
KIC 9468611     &                &  ROT          & 13.144  & 2.193 &    18 & 15 &  1 &  290 & 263  & 11710 &  1 & 1.98 & B9IV \\
KIC 6128830     & BD+41 3394     &  SXARI        &  9.190  & 0.206 &  1526 & 98 &  1 &   56 &  15  & 12600 & 23 & 2.82 & B6pHgMn \\
KIC 7974841     & HD 187139      &  ROT          &  8.160  & 0.255 &    68 & 23 &  2 &   22 &  33  & 10650 & 24 & 1.55 & B8V \\
KIC 5477601     &                &  ROT          & 12.793  & 0.192 &   374 & 29 &  1 &   25 &  88  & 11710 &  1 & 2.12 & B9V \\
KIC 5130305     & HD 226700      &  ROT          & 10.210  & 2.151 &    11 & 15 &  2 &  330 & 155  & 10670 & 23 & 2.03 & B9 \\
KIC 8324268     & HD 189160      &  SXARI(ACV:)  &  7.900  & 0.498 & 11890 & 21 &  2 &   72 &  31  & 11710 &  1 & 2.09 & B9pSiCr \\
KIC 8389948     & HD 189159      &  ROT          &  9.140  & 0.994 &    12 & 20 &  2 &  158 & 142  & 10240 & 23 & 1.99 & B9.5IV-V\\
KIC 5479821     & HD 226795      &  ROT          &  9.890  & 0.588 & 13586 & 25 &  2 &  119 &  85  & 14810 & 23 & 2.84 & B8 \\
EPIC 206326769  & HD 211838      &  SXARI        &  5.346  & 0.893 &   105 & 27 &  1 &  314 &  66  & 13520 &  1 & 3.16 & B8IIIp:(MnHg?)\\
EPIC 206097719  & HD 213781      &  SXARI        &  9.000  & 0.181 &   456 & 24 &  0 &   62 &  34  & 15330 &  1 & 3.36 & B7Si \\
\hline
\end{tabular}
}
\end{center}
\end{table*}

\section{Rotational modulation}

Since there is evidence that rotational modulation is present in nearly half 
of the {\it Kepler} A stars \citep{Balona2013c, Balona2017a}, it is 
reasonable to presume that rotational modulation may also be common among B 
stars.  The physics of the outer layers of these stars are very similar and we 
expect continuity in the properties of early A stars and late B stars.

Low frequencies similar to those expected from rotational modulation can
also be found in  mid- to late-B stars.  These are the SPB variables which
pulsate with multiple frequencies in the range 0.3--3.0\,d$^{-1}$.  For a late 
B star with $v_e =$50--100\,km\,s$^{-1}$, we expect $\nu_{\rm rot} \approx 
0.3$\,d$^{-1}$ which is at the low frequency end of the SPB range.  Therefore 
if a SPB frequency peak is mistaken for rotation, it most probably will be at 
a higher frequency.  This will result in a $v_e$ larger than $v\sin i$ and
will thus appear to confirm rotational modulation.

A single low frequencies or a low frequency and its harmonic could possibly
arise from SPB pulsation simply by coincidence.  There is no known mechanism
which preferentially selects just a single mode or a mode and its harmonic.
Unless evidence for such a selection mechanism is found, it is difficult to
accept SPB as a possible explanation for more than a few stars. In any case, 
pulsation can only occur if the star is within the instability strip.  While 
the SPB instability strip  may be extended to cooler stars by the effect of 
rotation, this explanation will ultimately fail for sufficiently cool stars.  
There are plenty of hot A stars which show a single low frequency or low 
frequency and harmonic well outside the SPB instability strip, no matter how 
much it can reasonably be extended.  These cannot be due to SPB pulsation.  It 
is thus reasonable to assume that a single peak or a peak and its harmonic, for
which the evidence indicates rotational modulation in A stars, is also a result
of rotational modulation in the B stars.

It is also possible that low frequencies may be a result of binarity or
Doppler beaming.  Considering the fact that most of the observed frequencies 
are around 1\,d$^{-1}$, it follows that the components must be rather close to 
each other.  Under these conditions we may expect to see eclipses or partial 
eclipses in most of the stars.  It is for this reason that we tend to assign a 
classification of EB or EA (rather than ROT) to stars with amplitudes in 
excess of a few parts per thousand.  It turns out that the large number of B 
stars classified as ROT all have small amplitudes (typically around 150 ppt).
The binarity explanation therefore requires that all these stars have grazing 
eclipses,which is very unlikely.  

For the reasons discussed above, we are confident that a ROT classification 
does indeed have a high probability of being due to rotational modulation,
but do not exclude some contamination from  pulsation and binarity.  Our aim
is not to prove that the ROT stars are due to rotational modulation, but
merely to show that rotational modulation is not excluded.  

The frequencies listed as $\nu_{\rm rot}$ in Tables\,\ref{bstars} and 
\ref{add} are all highly significant according to the false alarm
probability \citep{Scargle1982}.  Probabilities that the specified frequency
is due to noise are always less than $10^{-6}$ and the ratio of the peak 
amplitude to background noise level is always greater than 10.  The typical 
peak amplitude is around 135\,ppm.  

\begin{figure}
\centering
\includegraphics[]{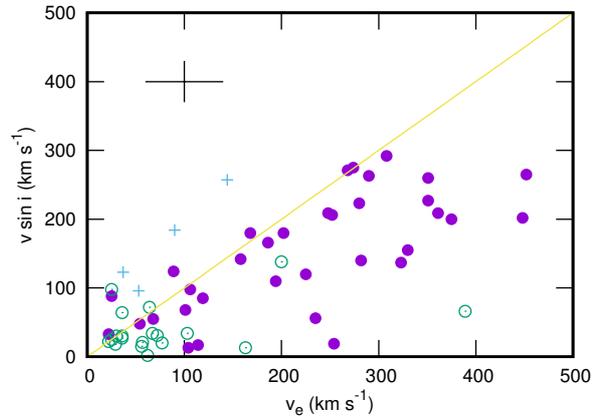}
\caption{The projected rotational velocity, $v\sin i$, as a function of
equatorial rotational velocity, $v_e$, for stars in the {\it TESS, Kepler} 
and {\it K2} fields (filled circles).  The open circles are SXARI stars which 
are rotational variables due to chemical abundance variations across the 
surface.  The small crosses are discrepant stars discussed in the text.  The 
straight line has unit slope, corresponding to $\sin i = 1$.  The cross at the 
top left is the approximate 1-$\sigma$ error bars.}
\label{keprot}
\end{figure}

As already mentioned, one test for rotational modulation is to compare the
equatorial rotational velocity, $v_e$, obtained from $\nu_{\rm rot}$ and the
stellar radius,  with $v\sin i$.  For this purpose, values of $v\sin i$ in 
Tables\,\ref{bstars} and \ref{add} were obtained from the catalogue of 
\citet{Glebocki2005}.  A few more recent values were found using
SIMBAD.  Fig.\,\ref{keprot} shows $v\sin i$ as a function of
$v_e$ for the {\it TESS} main sequence stars identified as ROT (solid circles)
or SXARI (open circles) in Tables\,\ref{bstars} and \ref{add}.  As expected, 
most stars fall below the $\sin i = 1$ line.  If the variation is not related 
to rotation, one would have expected both sides of the $\sin i = 1$ line to 
be populated.

The typical error in $v\sin i$ for B stars can be estimated from the
catalogue of \citet{Glebocki2005}.  The error increases with $v\sin i$ and
ranges between 0 and 60\,km\,s$^{-1}$.  A representative value of
$\sigma_{v\sin i} = 30$\,km\,s$^{-1}$ is reasonable.  From the error in
$\log{L/L_\odot}$ and $T_{\rm eff}$ it is easy to calculate the error in
$v_e$.  This error depends almost entirely on the error in $T_{\rm eff}$.
The contribution from the luminosity error is small while the contribution
from the error in $\nu_{\rm rot}$ is entirely negligible.  The typical value 
for the error in the derived equatorial rotational velocity is 
$\sigma_{v_e} \approx 40$\,km\,s$^{-1}$.  These error bars are shown in 
Fig.\,\ref{keprot}.  

The distribution of points in Fig.\,\ref{keprot} is roughly what one would
expect for rotational modulation.  Most of the stars would be seen
equator-on and hence most of the points will be near the $\sin i = 1$ line. 
Stars which are nearly pole-on would not show rotational modulation, which
is consistent with the lack of stars at the bottom right corner of the
figure.

An estimate for the angle of inclination of a particular star can be found
by dividing $v\sin i$ by $v_e$.  The error in the resulting 
value of $\sin i$ clearly increases as $v_e$ approaches zero
because $\sin i$ is the ratio of two small numbers and even a small 
error in $v\sin i$ and/or $v_e$ will lead to a large error in
$\sin i$.  Only rapidly rotating stars are useful in estimating 
$\sin i$.  As $v_e$ approaches zero, the true range of
$v\sin i$ also tends to zero and an increasing precision in 
$v\sin i$ is required to prevent it from exceeding $v_e$ due to 
measurement errors.  Since the error in $v\sin i$ is roughly the same
over the whole range of $v_e$, it is inevitable that an increasing
number of stars will lie above the $\sin i = 1$ line as $v_e$ tends
to zero.

Several stars lie above the $\sin i = 1$ line, including three SXARI 
stars for which rotational modulation is universally accepted as the cause
of the light variation.  TIC\,270070443 is only slightly discrepant and a 
small change in $v\sin i$ or $v_e$ will bring it below the 
$v\sin i = 1$ line.  The $v\sin i$  values of TIC\,280051467 
range from 36--64\,km\,s$^{-1}$ and adopting the smaller value will 
remove the discrepancy.  The same is true for the most discrepant SXARI star, 
TIC\,204095429 with $v\sin i$ in the range 18--98\,km\,s$^{-1}$.  

There is, of course, nothing unphysical about stars lying above the 
$v\sin i = 1$ line.  The location of a star in this diagram depends on 
two quantities which are both subject to error.  A star lying above the line
by more than three standard deviations is probably a result of
mis-classification.  Stars that do lie above the line  may simply be a result 
of an error in $v\sin i$ and/or $v_e$.  The error in $v_e$, 
for example, is sensitive to the error in $T_{\rm eff}$.  The 
interpretation of $v\sin i > 1$ in terms of the angle of inclination is 
that the star may be mis-classified as ROT or that the errors are too large for
$i$ to be measured, but that the inclination is probably nearly 
equator-on. 

The most discrepant ROT star is TIC\,355477670 with $v\sin i =
123$\,km\,s$^{-1}$ \citep{Zorec2012}, $v_e = 37$\,km\,s$^{-1}$. The 
estimated $v_e$ is three times smaller than $v\sin i$.  
This star has the lowest amplitude (21\,ppm) among the {\it TESS} 
stars, though the periodogram peak at 0.272\,d$^{-1}$ is prominent. 
{\it KELT} ground-based photometry gives a rotation period 
$P_{\rm rot} = 27.15180$\,d \citep{Oelkers2018}, which would make the 
discrepancy still larger.  No evidence of such a  period is found in the 
{\it TESS} data.  Perhaps the observed periodicity is of a binary nature.

TIC\,259862349 with $v\sin i = 96$\,km\,s$^{-1}$ and 
$v_e = 53$\,km\,s$^{-1}$ appears to be surrounded by a debris-disk
\citep{Welsh2018}.  This is a another low-amplitude star (51\,ppm) with a
strange light curve.  There is sudden doubling in the light amplitude
variation which looks to be of instrumental origin.  Although the
periodogram peak is sharp, it is surrounded by noise which is likely related
to the sudden change in the appearance of the light curve. 

EPIC\,202061205 is a variable of unknown type with a period of 3.414701\,d
\citep{Watson2006}, which is the same as in Table\,\ref{add}. 
The variability could possibly be a binarity effect considering that the 
amplitude is fairly large.

EPIC\,205417334 has a period of 1.0877\,d according to
\citet{Rebull2018}, the same as in Table\,\ref{add}, but no 
variability classification is given.  Perhaps this is another binary.

It should be noted that the ROT classification was made with no
prior knowledge of the literature on these stars.  However, one star 
(EPIC 202909059) was  removed after it was found to be a spectroscopic
binary (it was the most discrepant point in the $v\sin i$/$v_e$
diagram).  Thus there is no reason to suspect a bias regarding the 
allocation of the ROT class.  Only four or five ROT stars are found to be
somewhat discrepant which suggests that there are probably not many
mis-classifications.

\section{Notes on some ROT stars}

Stars which lie below the $\sin i = 1$ line but that are of interest are
discussed below.

TIC\,38602305 (HD\,27657) is an optical double with a separation of 4\,arcsecs and 
a magnitude difference  $\Delta V = 1.6$\,mag with spectral types B9III and 
B9V.  A period of 27.15\,d was reported by \citet{Oelkers2018} from 
{\it KELT} ground-based photometry. The {\it TESS} periodogram shows a peak 
with period 2.976\,d and three of its harmonics (Fig.\,\ref{phase}).  

TIC\,47296054 (HD\,214748) is a classical Be star.  Three harmonics of the
fundamental at 0.836\,d$^{-1}$ are clearly visible.  The star seems to be a
typical ROT variable.  In addition,  there is an anomalous peak at 
0.432\,d$^{-1}$ which is not part of the harmonic sequence. Thus a SPB+ROT
classification was assigned.

TIC\,139468902 (HD\,213155) has a rotation period of 5.97\,d from {\it KELT}
photometry \citep{Oelkers2018}.  The {\it TESS} data, on the other hand,
has a fundamental period of 0.455\,d with the first harmonic having a similar 
amplitude.  No other significant frequencies seem to be present.

TIC\,150357404 (HD\,45796) has a rotation period of 2.78\,d from {\it KELT}
photometry \citep{Oelkers2018}.  The {\it TESS} data gives $P_{\rm rot} =
1.56$\,d.  The periodogram shows two weak sidelobes surrounding the central
peak, hence the SPB+ROT? classification.  The uncertainty is that perhaps
the main peak is a pulsation mode and not due to rotation.

TIC\,152283270 (HD\,208433).  This is a visual double separated by
0.6\,arcsecs with $\Delta V = 2.26$\,mag.  The periodogram shows a single peak 
at 0.434\,d$^{-1}$ assumed to be rotation, but the light curve also shows a 
single clear eclipse with a duration of about 0.6\,d and a depth of about 
1\,millimag.  Hence the ROT?+EA classification.  The uncertainty is due to the
relatively low S/N of the supposedly rotation peak and the absence of
harmonics.

TIC\,238194921 (HD\,24579). Apart from the peak at $\nu_{\rm rot} =
0.727$\,d$^{-1}$ and its harmonic, the are two additional peaks, the one at
1.812\,d$^{-1}$ having the largest amplitude (190\,ppm).  The other peak is 
close to $\nu_{\rm rot}$ and with similar amplitude.  These additional peaks 
indicate that this may be an SPB star.  Hence the classification SPB+ROT.

TIC\,262815962 (HD\,218976) is a visual double with separation of
1.6\,arcsecs and $\Delta V = 3.91$\,mag.  A rotation period of 27.15\,d is
reported by \citet{Oelkers2018} from {\it KELT} data.  The {\it TESS} data give
$P_{\rm rot} = 2.71$\,d.

TIC\,271503441 (HD\,2884) is a visual double with separation of
2.5\,arcsecs and $\Delta V = 9.1$\.mag.  In fact, there are another five
associated, more widely separated, stars. The brightest star, $\beta^1$~Tuc, 
does not seem to be a spectroscopic binary \citep{Chini2012}.  The {\it TESS} 
data show a peak at 3.056\,d$^{-1}$ and its harmonic and nothing else of
significance.

TIC\,271971626 (HD\,62153) is a visual double separated by 1.9\,arcsecs
and of equal brightness and spectral types.  The main periodogram peak at
0.215\,d$^{-1}$ is assumed to be due to rotation.  There is also a peak of
lower amplitude ($A = 30$\,ppm) at 5.820\,d$^{-1}$ and its first harmonic 
($A = 13$\,ppm).  The high frequency must be a pulsation, so we have
assigned the MAIA class as the star lies between the red edge of the
$\beta$~Cep variables and the blue edge of the $\delta$~Sct stars.

TIC\,277103567 (HD\,37935) is a classical Be star. The periodogram shows a very
simple series of peaks with fundamental at 1.496\,d$^{-1}$ and up to three
harmonics.  The first harmonic has the highest amplitude. This seems to be a
typical ROT variable.

TIC\,279430029 (HD\,53048) is another classical Be star.  The periodogram 
is very simple and consists of just two peaks: the fundamental at
1.784\,d$^{-1}$ and the first harmonic which has twice the amplitude.  This
is a ROT variable.

KIC\,293268667 (HD\,47478).  The periodogram shows multiple peaks in the
range 2.8--3.9\,d$^{-1}$ and also at 6.6--7.1\,d$^{-1}$.  The multiple peaks
suggest a SPB star.  The harmonic of the main peak at 3.390\,d$^{-1}$ is 
visible with the same amplitude as the fundamental.  Hence the
classification SPB+ROT.

TIC\,300010961 (HD\,55478).  The periodogram shows a strong peak at
0.683\,d$^{-1}$ with three visible harmonics.  In addition there are multiple 
peaks in the range 17-45\,d$^{-1}$.  Because of its location in the H-R diagram 
we classify it as a MAIA variable.  The star is a visual double (2.2\,arcsecs, 
$\Delta V = 2.9$\,mag).

TIC\,300329728 (HD\,59426).  The fundamental peak has a large amplitude.
For this reason it might be considered an ellipsoidal (ELL) variable, but could
also be ROT.  The first harmonic is prominent.

TIC\,308537791 (HD\,67277).  The fundamental at 0.527\,d$^{-1}$ and its
first harmonic is present.  In addition, a peak at 7.397\,d$^{-1}$ and at 
least three of its harmonics can be seen.  The high frequency suggests
pulsation, so it has been given the classification ROT+MAIA.

TIC\,355653322 (HD\,224686) is a classical Be star.  The periodogram shows
only a weak peak at 1.266\,d$^{-1}$ and its first harmonic.  Apart from
this, no other significant peaks are present. 

TIC\,419065817 (HD\,1256). The periodogram shows a peak at 1.474\,d$^{-1}$ and 
at least three harmonics.  However, there is also an unrelated  peak at 
0.226\,d$^{-1}$.

TIC\,441182258 (HD\,210934) has a weak peak in the periodogram at 
1.274\,d$^{-1}$ and some unresolved structure around 0.5\,d$^{-1}$.  This
is classified as SPB+ROT?

EPIC\,211116936 (HD\.23324) has a strong frequency peak consisting of two close 
components.  The first harmonic of one of the components is present, from which
we suggest a ROT classification, but other low-frequency peaks exist which 
indicate  that it is also SPB.  The star is a member of the Pleiades.

KIC\,9278405 (ILF1 +45 284) was classified as SPB by \citet{McNamara2012}
The star has a strong peak with multiple close components and a weak first 
harmonic.  It could be indeed classified as SPB, but differential rotation will
also account for the structure in the main peak.  

KIC\,5479821 (HD\,226795) has a single sharp peak and harmonic. \citet{McNamara2012}
classified the star as either a binary or rotational variable with the period 
listed in Table\,\ref{add}.

\section{Flare stars}

Optical stellar flares are usually associated with active M dwarfs which can 
dramatically increase in brightness over a broad wavelength range from 
X-rays to radio waves for anywhere from a few minutes to a few hours. The 
rapid rise in brightness is followed by a slow decay with a time-scale from
minutes to hours. The largest change in brightness occurs at short
wavelengths: a rise of one magnitude in the V band is typically
accompanied by a rise of five magnitudes in the U band.  The amplitudes are
much lower in the very wide band used in the {\it TESS} and {\it Kepler}
observations.

Flares in the Sun are caused by energy released by the re-connection of 
magnetic field lines in the outer atmosphere.  The energies released in solar
flares are in the range $10^{29}$--$10^{32}$\,ergs.  Flares in active M dwarfs 
are  typically 10--1000 times more energetic than solar flares.  Although we 
have a very limited understanding of stellar flares, it is thought that the
underlying mechanism is essentially the same as in the Sun.

The {\it Kepler} mission has resulted in the discovery of ``superflares''
in solar-type stars with energies in the range $10^{33}$--$10^{36}$\,ergs
\citep{Maehara2012}.  More surprisingly, flaring was found in about
2.5\,percent of A stars \citep{Balona2012c, Balona2013c,Balona2015a}. 
Spectroscopic observations of the A-type flare stars suggest significant
contamination by field stars in the {\it Kepler} aperture and that many of
the stars are spectroscopic binaries \citep{Pedersen2017}.  The possibility
that the flares originate in a companion certainly cannot be discounted. 
However, it should be noted that the flares associated with A stars are
100--1000 times more energetic than those in typical M or K dwarfs
\citep{Balona2015a}, suggesting that they may originate in the A star
and not on a late-type companion.

\begin{figure}
\centering
\includegraphics[]{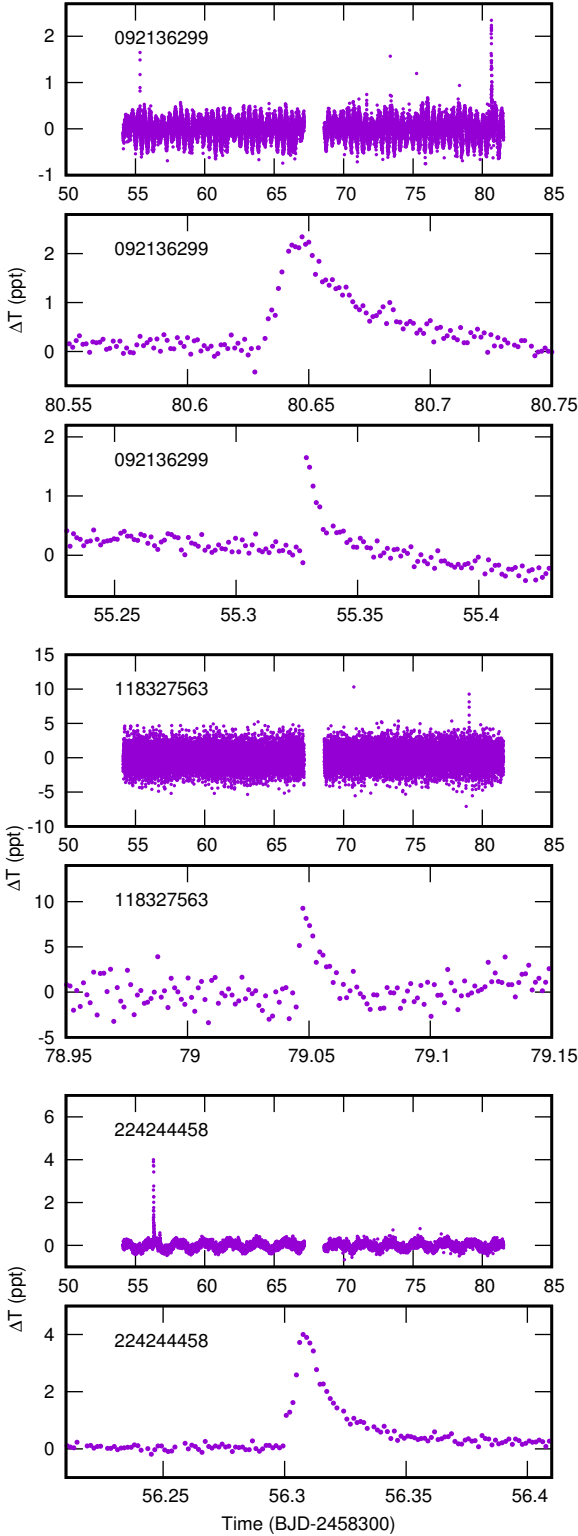}
\caption{Light curves of three stars (TIC numbers given) which appear to
show one or more flares.  The units for the light variation are parts per
thousand (ppt).}
\label{flare}
\end{figure}

In the course of inspection of the {\it TESS} light curves, we came across three
stars which appear to flare (Fig.\,\ref{flare}).  TIC\,92136299 is a normal
B9.5IV star with clear periodic variations suggestive of rotational modulation.
The periodogram shows a main peak at 2.251\,d$^{-1}$  and its first harmonic.
In addition, there is a third peak at 2.642\,d$^{-1}$ which could be 
interpreted as a second ``spot'' in a differentially-rotating star. Two 
flare-like events can be seen which have the typical sharp rise and 
slow decay of a stellar flare. According to \citet{Chini2012}, the star is 
not a spectroscopic binary.

TIC\,92136299 ($\omega^2$\,Aqr A) has a close companion, ($\omega^2$\,Aqr B)
with spectral type A5IVpec with the H\&K CaII lines in emission
\citep{Gahm1983}.   The stars are separated by 5\,arcsec and magnitude 
difference of 6\,mag.  The pair is an X-ray source \citep{Marakov2003}.

A flare is also found in TIC\,118327563 which is a subdwarf B star.  
\citet{Bagnulo2015} was unable to detect a significant magnetic field.  The 
rapid light variation of 4.372\,d$^{-1}$ is clearly seen in the light curve.  
The periodogram shows two closely-spaced peaks at 4.372 and 4.203\,d$^{-1}$ 
and their first harmonics. 

The flare on TIC\,224244458 ($\beta$\,Scl) is very interesting as this is
a HgMn star.  The HgMn stars are chemically peculiar stars containing an 
excess of P, Mn, Ga, Sr, Yt, Zr, Pt and Hg.  They lack a strong dipole 
magnetic field and are slow rotators.  For $\beta$\,Scl, \citet{Bychkov2009} 
quotes a (null) magnetic field measurement of $61 \pm 36$\,G.  
\citet{Chini2012} found that the star is not a spectroscopic binary.

We can calculate the approximate flare energy from the area occupied by
the flare in the light curve and the stellar luminosity.  These turn out to 
be approximately $10^{36}$\,ergs for TIC\,92136299 and TIC\,224244458 and 
$10^{35}$\,ergs for TIC\,118327563, which are all considerably larger than the
most energetic flares in typical K or M dwarfs, but similar to the flare 
energies in A stars.  This does not prove that the flares originate on 
B stars, but is an indication that this might be the case. 

\section{Other stars of interest}

The only $\beta$~Cep star in the sample is TIC\,69925250 (HN\,Aqr). 
{\it TESS} observations of the star are discussed by \citet{Handler2019}.

There are 25 stars classified as SPB variables in the {\it TESS} data. The 
classification is based on the presence of multiple low-frequency peaks.  In 
some stars one may interpret a strong peak and its harmonic as possibly due 
to rotation.  These have been designated SPB+ROT variables and are treated as 
both SPB and rotational variables.

The 12 stars considered to be classical Be stars are designated in the last
column of Table\,\ref{bstars} by ``(Be)''.  Many of them display broad
multiple peaks considered to be a result of nonradial pulsation or variable
circumstellar obscuration.  These are classified as BE variables, in accordance 
with the GVCS definition.  In addition to the broad peaks, TIC\,270219259 also 
shows a strong, sharp peak at 7.445\,d$^{-1}$.  The star is too cool for a BCEP, 
so we have classified it as BE/MAIA.  TIC\,308748912 does not have any 
significant peaks with frequencies higher than 0.5\,d$^{-1}$ and with 
amplitudes above 10\,ppm.  Four classical Be stars have already been
discussed in Section 8 and appear to be normal rotational variables.  
Two additional classical Be stars, EPIC\,210788932 and EPIC\,217692814 
also seem to be typical ROT variables.

Of the 7 stars classified as MAIA variables, three have $v\sin i$
measurements.  This brings to 16 the number of suspected MAIA stars with
known $v\sin i$, for which the mean is $<v\sin i> = 111 \pm
23$\,km\,s$^{-1}$.  This does not suggest that rapid rotation is an
important factor, unlike the FaRPB stars. 

TIC\,169285097 (J2344-3427) is a known subdwarf pulsating B star 
\citep{Holdsworth2017}. The {\it TESS} data show that it pulsates in both 
long- and short periods (hybrid V361\,Hya/V1093\,Her star).  There are dozens 
of peaks in the range $8 < \nu < 60$\,d$^{-1}$, but also a few peaks in the 
range $220 < \nu < 250$\,d$^{-1}$.  The other sdB stars and the white dwarf 
TIC\,31674330 appear to be constant.

Another class of interest are rotational variables with detected magnetic 
fields in the first {\it TESS} data release.  These are being investigated by
David-Uraz et al. (2019, in preparation).

\section{Conclusions}

For the last half-century the view that stars with radiative envelopes 
cannot sustain a magnetic field and are therefore devoid of starspots has 
been generally assumed.  Suspicions that this was not the case have 
occasionally arisen, particularly from inspection of good ground-based 
photometric time series of $\delta$~Sct and $\beta$~Cep stars, where the 
presence of significant low frequencies has sometimes been noted.
However, owing to the fact that the rotational periods of A and B stars are
close to one day, it is very difficult to detect low-amplitude rotational
modulation of A and B stars in ground-based photometry from a single site.

With the advent of precise time-series photometry from space it has become
apparent that rotational modulation is very likely present in about half the 
A stars in the {\it Kepler} field \citep{Balona2013c, Balona2017a}.  This is
demonstrated by the fact that the photometric period distribution closely
matches the expected distribution from spectroscopic $v\sin i$ measurements
\citep{Balona2013c}.  Furthermore, the expected relation between the
equatorial rotational velocity, $v_e$, estimated from the photometric period 
and $v\sin i$ is also present \citep{Balona2017a}.

There are very few B stars in the {\it Kepler} and {\it K2} fields (most of
them are late B stars), too few to calculate the $v_e$ distribution and 
compare it with the $v_e$ distribution expected from spectroscopic $v\sin i$
measurements from field stars in the same temperature range.  
\citet{Balona2016a} was able to identify presumed rotational modulation in 
many B stars observed with the {\it K2} mission, but for the most part 
$v\sin i$ measurements were not available to test the relationship with $v_e$.

In this paper we examined the light curves of 160 B stars observed by {\it TESS}
and classified them according to variability type. It appears that a large 
fraction of these stars may be rotational variables without any known chemical 
peculiarity.  \citet{Balona2016a} had already arrived at the same conclusion 
from a sample of {\it K2} B stars.

Using {\it Gaia} parallaxes, the luminosities of these main sequence stars were
estimated, from which the radii can be found.  From the radii and the
photometric periods, the equatorial rotational velocities can be  determined.
The expected relationship between projected rotational velocity and the 
equatorial rotational velocity is found, confirming that the photometric 
periods are consistent with rotation.  

Out of the 112 main-sequence B stars observed by {\it TESS}, 45 were 
classified as rotational variables or possible rotational variables. This 
fraction (about 40\,percent) is similar to the fraction of rotational 
variables found among the A stars \citep{Balona2013c} and a confirmation of a 
similar result from B stars in the {\it K2} field \citep{Balona2016a}.  
Rotational variability appears to be the most common type of light variation 
among A and B main sequence stars.  None of these variables are known to be 
chemically peculiar.

This result calls into question current models of the outer layer of stars 
with radiative envelopes.  \citet{Cantiello2009} and \citet{Cantiello2011} 
have suggested that magnetic fields produced in subsurface convection zones 
could appear on the surface. Thus localized magnetic fields could be 
widespread in those early type stars with subsurface convection. Magnetic 
spots of size comparable to the local pressure scale height are predicted
to manifest themselves as hot, bright spots \citep{Cantiello2011}.   Recent
observations from space indicate that bright spots may have been detected in
some O stars \citep{Ramiaramanantsoa2014, David-Uraz2017,
Ramiaramanantsoa2018}.  It is also possible that differential rotation in the 
A and B stars may be sufficient to create a local magnetic field via dynamo 
action \citep{Spruit1999, Spruit2002, Maeder2004}.  Whether or not any of these 
ideas relates to rotational modulation as observed in A and B stars requires 
further work.

\section*{Acknowledgments}

LAB wishes to thank the National Research Foundation of South Africa for 
financial support.  GMM acknowledges funding by the STFC consolidated grant 
ST/R000603/1.  GH and SC gratefully acknowledge funding through grant
2015/18/A/ST9/00578 of the Polish National Science Centre (NCN). GAW
acknowledges Discovery Grant support from the Natural Sciences and
Engineering Research Council (NSERC) of Canada.  ADU acknowledges the support 
of the Natural Science and Engineering Research Council of Canada (NSERC).

This paper includes data collected by the {\it TESS} mission. Funding for the 
{\it TESS} mission is provided by the NASA Explorer Program. Funding for the 
{\it TESS} Asteroseismic Science Operations Centre is provided by the Danish 
National Research Foundation (Grant agreement no.: DNRF106), ESA PRODEX
(PEA 4000119301) and Stellar Astrophysics Centre (SAC) at Aarhus University. 
We thank the {\it TESS} and TASC/TASOC teams for their support of the present
work.

This research has made use of the SIMBAD database, operated at CDS, 
Strasbourg, France.

This work has made use of data from the European Space Agency (ESA) mission
{\it Gaia} (\url{https://www.cosmos.esa.int/gaia}), processed by the {\it
Gaia} Data Processing and Analysis Consortium (DPAC,
\url{https://www.cosmos.esa.int/web/gaia/dpac/consortium}). Funding for the
DPAC has been provided by national institutions, in particular the 
institutions participating in the {\it Gaia} Multilateral Agreement.

The data presented in this paper were obtained from the Mikulski Archive for 
Space Telescopes (MAST).  STScI is operated by the Association of Universities
for Research in Astronomy, Inc., under NASA contract NAS5-2655.

\bibliographystyle{mn2e}
\bibliography{rotn}

\begin{thebibliography}{82}
\expandafter\ifx\csname natexlab\endcsname\relax\def\natexlab#1{#1}\fi

\bibitem[{{Bagnulo} {et~al.}(2015){Bagnulo}, {Fossati}, {Landstreet}, \&
  {Izzo}}]{Bagnulo2015}
{Bagnulo} S., {Fossati} L., {Landstreet} J.~D., {Izzo} C., 2015, \aap, 583,
  A115

\bibitem[{{Balona}(1994)}]{Balona1994a}
{Balona} L.~A., 1994, \mnras, 268, 119

\bibitem[{{Balona}(2012)}]{Balona2012c}
---, 2012, \mnras, 423, 3420

\bibitem[{{Balona}(2013)}]{Balona2013c}
---, 2013, \mnras, 431, 2240

\bibitem[{{Balona}(2015)}]{Balona2015a}
---, 2015, \mnras, 447, 2714

\bibitem[{{Balona}(2016)}]{Balona2016a}
---, 2016, \mnras, 457, 3724

\bibitem[{{Balona}(2017)}]{Balona2017a}
---, 2017, \mnras, 467, 1830

\bibitem[{{Balona}(2018)}]{Balona2018c}
---, 2018, \mnras, 479, 183

\bibitem[{{Balona} {et~al.}(2015){Balona}, {Baran},
  {Daszy{\'n}ska-Daszkiewicz}, \& {De Cat}}]{Balona2015c}
{Balona} L.~A., {Baran} A.~S., {Daszy{\'n}ska-Daszkiewicz} J., {De Cat} P.,
  2015, \mnras, 451, 1445

\bibitem[{{Balona} {et~al.}(2016){Balona}, {Engelbrecht}, {Joshi}, {Joshi},
  {Sharma}, {Semenko}, {Pandey}, {Chakradhari}, {Mkrtichian}, {Hema}, \&
  {Nemec}}]{Balona2016c}
{Balona} L.~A., {Engelbrecht} C.~A., {Joshi} Y.~C., {et~al.}, 2016, \mnras,
  460, 1318

\bibitem[{{Balona} {et~al.}(2011){Balona}, {Pigulski}, {Cat}, {Handler},
  {Guti{\'e}rrez-Soto}, {Engelbrecht}, {Frescura}, {Briquet}, {Cuypers},
  {Daszy{\'n}ska-Daszkiewicz}, {Degroote}, {Dukes}, {Garcia}, {Green}, {Heber},
  {Kawaler}, {Lehmann}, {Leroy}, {Molenda-{\.Z}aaowicz}, {Neiner}, {Noels},
  {Nuspl}, {{\O}stensen}, {Pricopi}, {Roxburgh}, {Salmon}, {Smith},
  {Su{\'a}rez}, {Suran}, {Szab{\'o}}, {Uytterhoeven}, {Christensen-Dalsgaard},
  {Kjeldsen}, {Caldwell}, {Girouard}, \& {Sanderfer}}]{Balona2011b}
{Balona} L.~A., {Pigulski} A., {Cat} P.~D., {et~al.}, 2011, \mnras, 413, 2403

\bibitem[{{Bertelli} {et~al.}(2008){Bertelli}, {Girardi}, {Marigo}, \&
  {Nasi}}]{Bertelli2008}
{Bertelli} G., {Girardi} L., {Marigo} P., {Nasi} E., 2008, \aap, 484, 815

\bibitem[{{Brown} {et~al.}(2011){Brown}, {Latham}, {Everett}, \&
  {Esquerdo}}]{Brown2011a}
{Brown} T.~M., {Latham} D.~W., {Everett} M.~E., {Esquerdo} G.~A., 2011, \aj,
  142, 112

\bibitem[{{Bychkov} {et~al.}(2009){Bychkov}, {Bychkova}, \&
  {Madej}}]{Bychkov2009}
{Bychkov} V.~D., {Bychkova} L.~V., {Madej} J., 2009, \mnras, 394, 1338

\bibitem[{{Cantiello} \& {Braithwaite}(2011)}]{Cantiello2011}
{Cantiello} M., {Braithwaite} J., 2011, \aap, 534, A140

\bibitem[{{Cantiello} {et~al.}(2009){Cantiello}, {Langer}, {Brott}, {de Koter},
  {Shore}, {Vink}, {Voegler}, {Lennon}, \& {Yoon}}]{Cantiello2009}
{Cantiello} M., {Langer} N., {Brott} I., {et~al.}, 2009, \aap, 499, 279

\bibitem[{{Chandler} {et~al.}(2016){Chandler}, {McDonald}, \&
  {Kane}}]{Chandler2016}
{Chandler} C.~O., {McDonald} I., {Kane} S.~R., 2016, \aj, 151, 59

\bibitem[{{Charbonneau}(2014)}]{Charbonneau2014}
{Charbonneau} P., 2014, \araa, 52, 251

\bibitem[{{Chini} {et~al.}(2012){Chini}, {Hoffmeister}, {Nasseri}, {Stahl}, \&
  {Zinnecker}}]{Chini2012}
{Chini} R., {Hoffmeister} V.~H., {Nasseri} A., {Stahl} O., {Zinnecker} H.,
  2012, \mnras, 424, 1925

\bibitem[{{Chowdhury} {et~al.}(2018){Chowdhury}, {Joshi}, {Engelbrecht}, {De
  Cat}, {Joshi}, \& {Paul}}]{Chowdhury2018}
{Chowdhury} S., {Joshi} S., {Engelbrecht} C.~A., {et~al.}, 2018, \apss, 363,
  260

\bibitem[{{Crawford}(1978)}]{Crawford1978}
{Crawford} D.~L., 1978, \aj, 83, 48

\bibitem[{{David} \& {Hillenbrand}(2015)}]{David2015}
{David} T.~J., {Hillenbrand} L.~A., 2015, \apj, 804, 146

\bibitem[{{David-Uraz} {et~al.}(2017){David-Uraz}, {Owocki}, {Wade},
  {Sundqvist}, \& {Kee}}]{David-Uraz2017}
{David-Uraz} A., {Owocki} S.~P., {Wade} G.~A., {Sundqvist} J.~O., {Kee} N.~D.,
  2017, \mnras, 470, 3672

\bibitem[{{De Cat} \& {Aerts}(2002)}]{DeCat2002}
{De Cat} P., {Aerts} C., 2002, \aap, 393, 965

\bibitem[{{Dziembowski} {et~al.}(1993){Dziembowski}, {Moskalik}, \&
  {Pamyatnykh}}]{Dziembowski1993b}
{Dziembowski} W.~A., {Moskalik} P., {Pamyatnykh} A.~A., 1993, \mnras, 265, 588

\bibitem[{{Gahm} {et~al.}(1983){Gahm}, {Ahlin}, \& {Lindroos}}]{Gahm1983}
{Gahm} G.~F., {Ahlin} P., {Lindroos} K.~P., 1983, \aaps, 51, 143

\bibitem[{{Gaia Collaboration} {et~al.}(2018){Gaia Collaboration}, {Brown},
  {Vallenari}, {Prusti}, {de Bruijne}, {Babusiaux}, {Bailer-Jones}, {Biermann},
  {Evans}, {Eyer}, \& et~al.}]{Gaia2018}
{Gaia Collaboration}, {Brown} A.~G.~A., {Vallenari} A., {et~al.}, 2018, \aap,
  616, A1

\bibitem[{{Gaia Collaboration} {et~al.}(2016){Gaia Collaboration}, {Prusti},
  {de Bruijne}, {Brown}, {Vallenari}, {Babusiaux}, {Bailer-Jones}, {Bastian},
  {Biermann}, {Evans}, \& et~al.}]{Gaia2016}
{Gaia Collaboration}, {Prusti} T., {de Bruijne} J.~H.~J., {et~al.}, 2016, \aap,
  595, A1

\bibitem[{{Geier} {et~al.}(2017){Geier}, {{\O}stensen}, {Nemeth}, {Gentile
  Fusillo}, {G{\"a}nsicke}, {Telting}, {Green}, \& {Schaffenroth}}]{Geier2017}
{Geier} S., {{\O}stensen} R.~H., {Nemeth} P., {et~al.}, 2017, \aap, 600, A50

\bibitem[{{Gianninas} {et~al.}(2011){Gianninas}, {Bergeron}, \&
  {Ruiz}}]{Gianninas2011}
{Gianninas} A., {Bergeron} P., {Ruiz} M.~T., 2011, \apj, 743, 138

\bibitem[{{Glebocki} \& {Gnacinski}(2005)}]{Glebocki2005}
{Glebocki} R., {Gnacinski} P., 2005, VizieR Online Data Catalog, 3244

\bibitem[{{Gontcharov}(2017)}]{Gontcharov2017}
{Gontcharov} G.~A., 2017, Astronomy Letters, 43, 472

\bibitem[{{Gullikson} {et~al.}(2016){Gullikson}, {Kraus}, \&
  {Dodson-Robinson}}]{Gullikson2016}
{Gullikson} K., {Kraus} A., {Dodson-Robinson} S., 2016, \aj, 152, 40

\bibitem[{{Hall}(1972)}]{Hall1972}
{Hall} D.~S., 1972, \pasp, 84, 323

\bibitem[{{Handler} {et~al.}(2019){Handler}, {Pigulski},
  {Daszy{\'n}ska-Daszkiewicz}, {Irrgang}, {Kilkenny}, {Guo}, {Przybilla},
  {Alicavus}, {Kallinger}, {Pascual-Granado}, {Niemczura}, {Rozanski},
  {Chowdhury}, {Buzasi}, {Mirouh}, {Simon-Diaz}, {Moravveji}, \& {De
  Cat}}]{Handler2019}
{Handler} G., {Pigulski} A., {Daszy{\'n}ska-Daszkiewicz} J., {et~al.}, 2019,
  \apjl, {submitted}

\bibitem[{{Holdsworth} {et~al.}(2017){Holdsworth}, {{\O}stensen}, {Smalley}, \&
  {Telting}}]{Holdsworth2017}
{Holdsworth} D.~L., {{\O}stensen} R.~H., {Smalley} B., {Telting} J.~H., 2017,
  \mnras, 466, 5020

\bibitem[{{Huber} {et~al.}(2016){Huber}, {Bryson}, {Haas}, {Barclay},
  {Barentsen}, {Howell}, {Sharma}, {Stello}, \& {Thompson}}]{Huber2016}
{Huber} D., {Bryson} S.~T., {Haas} M.~R., {et~al.}, 2016, \apjs, 224, 2

\bibitem[{{Jenkins} {et~al.}(2016){Jenkins}, {Twicken}, {McCauliff},
  {Campbell}, {Sanderfer}, {Lung}, {Mansouri-Samani}, {Girouard}, {Tenenbaum},
  {Klaus}, {Smith}, {Caldwell}, {Chacon}, {Henze}, {Heiges}, {Latham},
  {Morgan}, {Swade}, {Rinehart}, \& {Vanderspek}}]{Jenkins2016}
{Jenkins} J.~M., {Twicken} J.~D., {McCauliff} S., {et~al.}, 2016, in \procspie,
  Vol. 9913, Software and Cyberinfrastructure for Astronomy IV, p. 99133E

\bibitem[{{Kron}(1947)}]{Kron1947}
{Kron} G.~E., 1947, \pasp, 59, 261

\bibitem[{{Maeder} \& {Meynet}(2004)}]{Maeder2004}
{Maeder} A., {Meynet} G., 2004, \aap, 422, 225

\bibitem[{{Maehara} {et~al.}(2012){Maehara}, {Shibayama}, {Notsu}, {Notsu},
  {Nagao}, {Kusaba}, {Honda}, {Nogami}, \& {Shibata}}]{Maehara2012}
{Maehara} H., {Shibayama} T., {Notsu} S., {et~al.}, 2012, \nat, 485, 478

\bibitem[{{Makarov}(2003)}]{Marakov2003}
{Makarov} V.~V., 2003, \aj, 126, 1996

\bibitem[{{Massey}(2002)}]{Massey2002}
{Massey} P., 2002, \apjs, 141, 81

\bibitem[{{McDonald} {et~al.}(2017){McDonald}, {Zijlstra}, \&
  {Watson}}]{McDonald2017}
{McDonald} I., {Zijlstra} A.~A., {Watson} R.~A., 2017, \mnras, 471, 770

\bibitem[{{McNamara} {et~al.}(2012){McNamara}, {Jackiewicz}, \&
  {McKeever}}]{McNamara2012}
{McNamara} B.~J., {Jackiewicz} J., {McKeever} J., 2012, \aj, 143, 101

\bibitem[{{McQuillan} {et~al.}(2013){McQuillan}, {Mazeh}, \&
  {Aigrain}}]{McQuillan2013}
{McQuillan} A., {Mazeh} T., {Aigrain} S., 2013, \apjl, 775, L11

\bibitem[{{McQuillan} {et~al.}(2014){McQuillan}, {Mazeh}, \&
  {Aigrain}}]{McQuillan2014}
---, 2014, \apjs, 211, 24

\bibitem[{{Michaud}(1970)}]{Michaud1970}
{Michaud} G., 1970, \apj, 160, 641

\bibitem[{{Miglio} {et~al.}(2007){Miglio}, {Montalb{\'a}n}, \&
  {Dupret}}]{Miglio2007}
{Miglio} A., {Montalb{\'a}n} J., {Dupret} M., 2007, Communications in
  Asteroseismology, 151, 48

\bibitem[{{Mowlavi} {et~al.}(2016){Mowlavi}, {Saesen}, {Semaan}, {Eggenberger},
  {Barblan}, {Eyer}, {Ekstr{\"o}m}, \& {Georgy}}]{Mowlavi2016}
{Mowlavi} N., {Saesen} S., {Semaan} T., {et~al.}, 2016, \aap, 595, L1

\bibitem[{{Nielsen} {et~al.}(2013){Nielsen}, {Gizon}, {Schunker}, \&
  {Karoff}}]{Nielsen2013}
{Nielsen} M.~B., {Gizon} L., {Schunker} H., {Karoff} C., 2013, \aap, 557, L10

\bibitem[{{Niemczura} {et~al.}(2015){Niemczura}, {Murphy}, {Smalley},
  {Uytterhoeven}, {Pigulski}, {Lehmann}, {Bowman}, {Catanzaro}, {van Aarle},
  {Bloemen}, {Briquet}, {De Cat}, {Drobek}, {Eyer}, {Gameiro}, {Gorlova},
  {Kami{\'n}ski}, {Lampens}, {Marcos-Arenal}, {P{\'a}pics}, {Vandenbussche},
  {Van Winckel}, {St{\c e}{\'s}licki}, \& {Fagas}}]{Niemczura2015}
{Niemczura} E., {Murphy} S.~J., {Smalley} B., {et~al.}, 2015, \mnras, 450, 2764

\bibitem[{{Oelkers} {et~al.}(2018){Oelkers}, {Rodriguez}, {Stassun}, {Pepper},
  {Somers}, {Kafka}, {Stevens}, {Beatty}, {Siverd}, {Lund}, {Kuhn}, {James}, \&
  {Gaudi}}]{Oelkers2018}
{Oelkers} R.~J., {Rodriguez} J.~E., {Stassun} K.~G., {et~al.}, 2018, \aj, 155,
  39

\bibitem[{{Oey} \& {Massey}(1995)}]{Oey1995}
{Oey} M.~S., {Massey} P., 1995, \apj, 452, 210

\bibitem[{{Paunzen} {et~al.}(2005){Paunzen}, {Schnell}, \&
  {Maitzen}}]{Paunzen2005}
{Paunzen} E., {Schnell} A., {Maitzen} H.~M., 2005, \aap, 444, 941

\bibitem[{{Paunzen} {et~al.}(2013){Paunzen}, {Wraight}, {Fossati}, {Netopil},
  {White}, \& {Bewsher}}]{Paunzen2013}
{Paunzen} E., {Wraight} K.~T., {Fossati} L., {et~al.}, 2013, \mnras, 429, 119

\bibitem[{{Pecaut} \& {Mamajek}(2013)}]{Pecaut2013}
{Pecaut} M.~J., {Mamajek} E.~E., 2013, \apjs, 208, 9

\bibitem[{{Pedersen} {et~al.}(2017){Pedersen}, {Antoci}, {Korhonen}, {White},
  {Jessen-Hansen}, {Lehtinen}, {Nikbakhsh}, \& {Viuho}}]{Pedersen2017}
{Pedersen} M.~G., {Antoci} V., {Korhonen} H., {et~al.}, 2017, \mnras, 466, 3060

\bibitem[{{Pedersen} {et~al.}(2019){Pedersen}, {Chowdhury}, {Johnston},
  {Bowman}, {Aerts}, {Handler}, {De Cat}, {Neiner}, {David-Uraz}, {Buzasi},
  {Tkachenko}, {Simon-Diaz}, {Moravveji}, {Sikora}, {Mirouh}, {Lovekin},
  {Cantiello}, {Daszy{\'n}ska-Daszkiewicz}, \& {Pigulski}}]{Pedersen2019}
{Pedersen} M.~G., {Chowdhury} S., {Johnston} C., {et~al.}, 2019, \apjl,
  submitted

\bibitem[{{Ramiaramanantsoa} {et~al.}(2014){Ramiaramanantsoa}, {Moffat},
  {Chen{\'e}}, {Richardson}, {Henrichs}, {Desforges}, {Antoci}, {Rowe},
  {Matthews}, {Kuschnig}, {Weiss}, {Sasselov}, {Rucinski}, \&
  {Guenther}}]{Ramiaramanantsoa2014}
{Ramiaramanantsoa} T., {Moffat} A.~F.~J., {Chen{\'e}} A.-N., {et~al.}, 2014,
  \mnras, 441, 910

\bibitem[{{Ramiaramanantsoa} {et~al.}(2018){Ramiaramanantsoa}, {Moffat},
  {Harmon}, {Ignace}, {St-Louis}, {Vanbeveren}, {Shenar}, {Pablo},
  {Richardson}, {Howarth}, {Stevens}, {Piaulet}, {St-Jean}, {Eversberg},
  {Pigulski}, {Popowicz}, {Kuschnig}, {Zoc{\l}o{\'n}ska}, {Buysschaert},
  {Handler}, {Weiss}, {Wade}, {Rucinski}, {Zwintz}, {Luckas}, {Heathcote},
  {Cacella}, {Powles}, {Locke}, {Bohlsen}, {Chen{\'e}}, {Miszalski}, {Waldron},
  {Kotze}, {Kotze}, \& {B{\"o}hm}}]{Ramiaramanantsoa2018}
{Ramiaramanantsoa} T., {Moffat} A.~F.~J., {Harmon} R., {et~al.}, 2018, \mnras,
  473, 5532

\bibitem[{{Rebull} {et~al.}(2018){Rebull}, {Stauffer}, {Cody}, {Hillenbrand},
  {David}, \& {Pinsonneault}}]{Rebull2018}
{Rebull} L.~M., {Stauffer} J.~R., {Cody} A.~M., {et~al.}, 2018, \aj, 155, 196

\bibitem[{{Reinhold} {et~al.}(2013){Reinhold}, {Reiners}, \&
  {Basri}}]{Reinhold2013}
{Reinhold} T., {Reiners} A., {Basri} G., 2013, \aap, 560, A4

\bibitem[{{Ricker} {et~al.}(2015){Ricker}, {Winn}, {Vanderspek}, {Latham},
  {Bakos}, {Bean}, {Berta-Thompson}, {Brown}, {Buchhave}, {Butler}, {Butler},
  {Chaplin}, {Charbonneau}, {Christensen-Dalsgaard}, {Clampin}, {Deming},
  {Doty}, {De Lee}, {Dressing}, {Dunham}, {Endl}, {Fressin}, {Ge}, {Henning},
  {Holman}, {Howard}, {Ida}, {Jenkins}, {Jernigan}, {Johnson}, {Kaltenegger},
  {Kawai}, {Kjeldsen}, {Laughlin}, {Levine}, {Lin}, {Lissauer}, {MacQueen},
  {Marcy}, {McCullough}, {Morton}, {Narita}, {Paegert}, {Palle}, {Pepe},
  {Pepper}, {Quirrenbach}, {Rinehart}, {Sasselov}, {Sato}, {Seager},
  {Sozzetti}, {Stassun}, {Sullivan}, {Szentgyorgyi}, {Torres}, {Udry}, \&
  {Villasenor}}]{Ricker2015}
{Ricker} G.~R., {Winn} J.~N., {Vanderspek} R., {et~al.}, 2015, Journal of
  Astronomical Telescopes, Instruments, and Systems, 1, 014003

\bibitem[{{Samus} {et~al.}(2009){Samus}, {Durlevich}, \& {et al.}}]{Samus2009}
{Samus} N.~N., {Durlevich} O.~V., {et al.}, 2009, VizieR Online Data Catalog,
  1, 2025

\bibitem[{{S{\'a}nchez-Bl{\'a}zquez} {et~al.}(2006){S{\'a}nchez-Bl{\'a}zquez},
  {Peletier}, {Jim{\'e}nez-Vicente}, {Cardiel}, {Cenarro},
  {Falc{\'o}n-Barroso}, {Gorgas}, {Selam}, \& {Vazdekis}}]{SanchezBlazquez2006}
{S{\'a}nchez-Bl{\'a}zquez} P., {Peletier} R.~F., {Jim{\'e}nez-Vicente} J.,
  {et~al.}, 2006, \mnras, 371, 703

\bibitem[{{Scargle}(1982)}]{Scargle1982}
{Scargle} J.~D., 1982, \apj, 263, 835

\bibitem[{{Silaj} \& {Landstreet}(2014)}]{Silaj2014}
{Silaj} J., {Landstreet} J.~D., 2014, \aap, 566, A132

\bibitem[{{Silva} \& {Napiwotzki}(2011)}]{Silva2011}
{Silva} M.~D.~V., {Napiwotzki} R., 2011, \mnras, 411, 2596

\bibitem[{{Soubiran} {et~al.}(2016){Soubiran}, {Le Campion}, {Brouillet}, \&
  {Chemin}}]{Soubiran2016}
{Soubiran} C., {Le Campion} J.-F., {Brouillet} N., {Chemin} L., 2016, \aap,
  591, A118

\bibitem[{{Spruit}(1999)}]{Spruit1999}
{Spruit} H.~C., 1999, \aap, 349, 189

\bibitem[{{Spruit}(2002)}]{Spruit2002}
---, 2002, \aap, 381, 923

\bibitem[{{Stassun} {et~al.}(2018){Stassun}, {Oelkers}, {Pepper}, {Paegert},
  {De Lee}, {Torres}, {Latham}, {Charpinet}, {Dressing}, {Huber}, {Kane},
  {L{\'e}pine}, {Mann}, {Muirhead}, {Rojas-Ayala}, {Silvotti}, {Fleming},
  {Levine}, \& {Plavchan}}]{Stassun2018}
{Stassun} K.~G., {Oelkers} R.~J., {Pepper} J., {et~al.}, 2018, \aj, 156, 102

\bibitem[{{Strassmeier}(2009)}]{Strassmeier2009}
{Strassmeier} K.~G., 2009, \aapr, 17, 251

\bibitem[{{Tkachenko} {et~al.}(2013){Tkachenko}, {Lehmann}, {Smalley}, \&
  {Uytterhoeven}}]{Tkachenko2013b}
{Tkachenko} A., {Lehmann} H., {Smalley} B., {Uytterhoeven} K., 2013, \mnras,
  431, 3685

\bibitem[{{Urbaneja} {et~al.}(2017){Urbaneja}, {Kudritzki}, {Gieren},
  {Pietrzy{\'n}ski}, {Bresolin}, \& {Przybilla}}]{Urbaneja2017}
{Urbaneja} M.~A., {Kudritzki} R.-P., {Gieren} W., {et~al.}, 2017, \aj, 154, 102

\bibitem[{{Watson} {et~al.}(2006){Watson}, {Henden}, \& {Price}}]{Watson2006}
{Watson} C.~L., {Henden} A.~A., {Price} A., 2006, Society for Astronomical
  Sciences Annual Symposium, 25, 47

\bibitem[{{Welsh} \& {Montgomery}(2018)}]{Welsh2018}
{Welsh} B.~Y., {Montgomery} S.~L., 2018, \mnras, 474, 1515

\bibitem[{{Wenger} {et~al.}(2000){Wenger}, {Ochsenbein}, {Egret}, {Dubois},
  {Bonnarel}, {Borde}, {Genova}, {Jasniewicz}, {Lalo{\"e}}, {Lesteven}, \&
  {Monier}}]{Wenger2000}
{Wenger} M., {Ochsenbein} F., {Egret} D., {et~al.}, 2000, \aaps, 143, 9

\bibitem[{{Wright} {et~al.}(2003){Wright}, {Egan}, {Kraemer}, \&
  {Price}}]{Wright2003}
{Wright} C.~O., {Egan} M.~P., {Kraemer} K.~E., {Price} S.~D., 2003, \aj, 125,
  359

\bibitem[{{Zorec} {et~al.}(2009){Zorec}, {Cidale}, {Arias}, {Fr{\'e}mat},
  {Muratore}, {Torres}, \& {Martayan}}]{Zorec2009}
{Zorec} J., {Cidale} L., {Arias} M.~L., {et~al.}, 2009, \aap, 501, 297

\bibitem[{{Zorec} \& {Royer}(2012)}]{Zorec2012}
{Zorec} J., {Royer} F., 2012, \aap, 537, A120

\end{thebibliography}

\label{lastpage}
\end{document}